\documentclass[epj]{svjour}
\usepackage{graphicx}
\usepackage[maxfloats=256]{morefloats}
\begin{document}
\title{Hadron production in elementary nucleon-nucleon reactions from low to ultra-relativistic energies}
\author{V. Kireyeu\inst{1}, I. Grishmanovskii\inst{2}, V. Kolesnikov\inst{1}, V. Voronyuk\inst{1},  and E. Bratkovskaya\inst{2,3}}
%
%
\institute{Joint Institute for Nuclear Research, Joliot-Curie 6, 141980 Dubna, Moscow region, Russia \and 
           Institute for Theoretical Physics, Johann Wolfgang Goethe Universit\"{a}t, Frankfurt am Main, Germany \and
           GSI Helmholtzzentrum f\"{u}r Schwerionenforschung GmbH, Planckstrasse 1, 64291 Darmstadt, Germany }
\date{Received: date / Revised version: date}
%
\abstract{
We study the hadron production in $p+p$, $p+n$ and $n+n$ reactions within 
the microscopic Parton-Hadron-Dynamics (PHSD) transport approach in comparison to
PYTHIA 8.2. We discuss the details of the "PHSD tune" of the Lund string model 
(realized by event generators FRITIOF and PYTHIA) in the vacuum 
(as in $N+N$ collisions) as well as its in-medium modifications relevant
for heavy-ion collisions where a hot and dense matter is produced.
We compare the results of PHSD and PYTHIA 8.2 (default version) for the excitation function 
of hadron multiplicities as well as differential rapidity $y$, transverse momentum $p_T$
and $x_F$ distributions in $p+p$, $p+n$ and $n+n$ reactions with the existing 
experimental data in the energy range $\sqrt{s_{NN}} = 2.7 - 7000$ GeV. 
We discuss the production mechanisms of hadrons and the role of final state interactions 
(FSI) due to the hadronic rescattering. We also show the influence of the possible 
quark-gluon plasma (QGP) formation on hadronic observables in $p+p$ collisions at LHC energies.
We stress the importance of developing a reliable event generator for elementary 
reactions from low to ultra-relativistic energies in view of actual and upcoming 
heavy-ion experiments.
\PACS{{ } elementary reactions, event generators}
} 

\titlerunning{Hadron production in elementary nucleon-nucleon reactions from low to ultra-relativistic energies}
\authorrunning{V. Kireyeu et al.}
\maketitle
\section{Introduction}
\label{intro}
An understanding of the mechanisms of multiparticle production in elementary 
nucleon-nucleon ($NN$) collisions in a wide energy range - from a few GeV up to a few TeV - 
is one of the challenging topics in hadron physics. This has a high 
impact on heavy-ion physics as well since in heavy-ion collisions (HIC)
one probes the matter created by many individual $NN$ scatterings - from primary
highly energetic $NN$ scattering during the initial phase of overlapping nuclei 
up to secondary low energy $NN$ collisions occurring during the final state 
interactions (FSI) of the expanding system.
Thus, for the description of the HIC one needs to know  the 
elementary hadron-hadron ($hh$) collisions: baryon-baryon ($BB$), meson-baryon ($mB$) and 
meson-meson ($mm$) collisions, in particular for hadron
multiplicities, i.e. flavour 'chemistry', as well as  their momentum 
distributions. Moreover, the elementary $NN$ reactions are used in HICs
as a "reference frame"  to study many physical effects related
to the properties of the hot and dense matter created in HICs.
For example, the most common way to present HIC results for hard probes
($i$ = charm or jets), is to show the ratio of their production probability in $A+A$
collisions relative to $p+p$ collisions scaled with the number of 
binary collisions $N_{bin}$: $R_{AA}^i=\sigma_{AA}^i/(\sigma_{pp}^i \cdot N_{bin})$.
The deviation of the ratio from unity provides information on the in-medium effects.

One of the most successful and commonly used models for the description
of elementary collisions from the GeV to the TeV energy range is 
the Lund string model \cite{LUND} which describes the energetic 
hadron-hadron collisions by the creation of excited color-singlet states, 
denoted by "strings", which are realized within the FRITIOF \cite{FRITIOF} 
and PYTHIA models \cite{Sjostrand:2006za} in terms of particle event generators.
A string is composed of two string ends corresponding to the leading constituent quarks 
(antiquarks) of the colliding hadrons and a color flux tube (color-electric field) in between. 
As the string ends recede, virtual $q\bar q$ or $qq\bar q\bar q$ pairs are produced in the
uniform color field by a tunneling process (described by the Schwinger formula \cite{Schwinger}), causing the breaking of the string and producing new matter
from field energy.

The Lund model is extremely successful in describing a huge variety of observables 
at high energies. The event generator PYTHIA is very often used by experimental
collaborations for a comparison with the measured data as well as for simulations 
of the detector set up.
The Lund model is employed in heavy-ion transport approaches for the
simulation of multiparticle production in elementary hadron-hadron 
collisions which happened during the time evolution of relativistic heavy-ion reactions.
The FRITIOF and PYTHIA event generators are incorporated in the off-shell 
Parton-Hadron-String Dynamics approach (PHSD) \cite{Cassing:2008sv,Cassing:2008nn,Cassing:2009vt,Bratkovskaya:2011wp,Linnyk:2015rco}
and it's early version Hadron-String Dynamics (HSD) \cite{Cassing:1999es}
(cf. the HSD review \cite{Cassing:1999es} for the description of string 
dynamics in HICs) as well as in the recent extension of the PHSD for 
the cluster formation, the Parton-Hadron-Quantum-Molecular Dynamics (PHQMD) 
\cite{Aichelin:2019tnk} based on the quantum-molecular dynamics (QMD) 
propagation of hadrons.
Moreover, PYTHIA is used in UrQMD \cite{Bass:1998ca,Bleicher:1999xi},
GiBUU \cite{GiBUU},  SMASH \cite{Weil:2016zrk} etc.
We note that there are alternative event generators for hadron-hadron
collisions such as  EPOS \cite{Werner:2005jf,EPOSLHC}, QGSJET \cite{QGSJET}, 
HERWIG \cite{HERWIG} etc.  

Most hadron-hadron event generators have been constructed for the description of 
ultra-relativistic $p+p$ collisions or cosmic rays at very high energies.
However, the use of $hh$ event generators in the transport approaches
for HICs has a very important specification: as mentioned above, the energy range of 
$hh$ reactions - taking place during the HIC evolution - is very wide, e.g. even 
if one considers $A+A$ collisions at LHC energies, the secondary reactions, 
which take place after the hadronization of the quark-gluon plasma created in such 
HIC collisions, cover a very broad interval of the invariant energy $\sqrt{s}$.
Thus, the $hh$ generator must have a wide range of applicability, i.e. from a
few GeV to a few TeV. In this respect the Lund event generators (FRITIOF and PYTHIA)
are quite suitable and provide a rather convincing description of inelastic $hh$ collisions
from high energies to the lower ones. However, some improvement of the model, i.e.
"tuning", is required for an extension to low energies: adjustment of the flavour chemistry
of the produced particles and their distributions. Moreover, in HICs the string fragments
in the hot and dense environment which might lead to a modification of the
fragmentation mechanism and the properties of the produced hadrons.
Such modifications have been incorporated in the FRITIOF 7.02 and PYTHIA 6.4 models
during the development of the PHSD(HSD) approach which we will call as "PHSD tune".

In this study we perform a systematic analysis of the the hadron  production 
in $p+p$, $p+n$ and $n+n$ reactions within the microscopic PHSD transport approach and 
PYTHIA 8.2 in its default version. We present the details of the "PHSD tune" of 
the Lund generators FRITIOF 7.02 
and PYTHIA 6.4 for the inelastic $hh$ collisions in the vacuum. Furthermore, we discuss 
the in-medium extension of the Lund string model for heavy-ion collisions where 
the string formation and decay occurs in a hot and dense environment. 
We provide a detailed comparison of the PHSD results with default (without any tunes) 
PYTHIA 8.2 \cite{PYTHIA82}  for $p+p$, $p+n$, $n+n$ collisions from a few GeV to a few TeV. 
We discuss the production mechanisms of hadrons and role of final state interactions 
due to the hadronic rescattering. We also show the influence of the possible quark-gluon plasma (QGP) formation
on hadronic observables for $p+p$ collisions at the LHC energy.

We stress that the necessity to develop a reliable event generator for 
the elementary reactions at low and intermediate energies is getting actual 
and timely since two new HIC accelerators -- 
the Facility for Antiproton and Ion Research (FAIR) in Darmstadt and 
the Nuclotron-based Ion Collider fAcility (NICA) in Dubna, 
will become operational in the next years and study nuclear matter at high 
baryon densities. Moreover, the presently running BES-II (Beam Energy Scan) experiments 
at RHIC, which includes a fixed target program, provide  experimental data in this 
energy regime.

Our paper is organized as follows:
after the Introduction we present the basic ideas of the PHSD approach in Section 2, 
then we step to the "PHSD tune" of the Lund model
in Section 3 and continue in Section 4 with the results for observables 
and the comparison of the PHSD and PYTHIA results with experimental data for 
inelastic $p+p$ collisions. 
We close our paper with Summary in Section 5.

\section{The PHSD~Approach}

We start with brief reminder of the basic ideas of the PHSD transport approach.
The~Parton--Hadron--String Dynamics  transport approach \cite{Cassing:2008sv,Cassing:2008nn,Cassing:2009vt,Bratkovskaya:2011wp,Linnyk:2015rco}
is a microscopic off-shell transport approach for the description of strongly interacting hadronic and partonic matter in and out-of equilibrium. It is based on the solution of
Kadanoff--Baym equations in first-order gradient expansion
\cite{Cassing:2008nn} employing `resummed' propagators from the dynamical
quasiparticle model (DQPM)~\cite{Cassing:2008nn,Peshier:2005pp}
for the partonic~phase. 

The DQPM provides an effective description
of the properties of the QGP in terms of strongly interacting quarks and gluons
with properties and interactions which are adjusted to reproduce lQCD results
on the thermodynamics of the equilibrated QGP at finite temperature $T$ and 
baryon (or quark) chemical potential $\mu_q$.  
Within the QGP phase, the partons (quarks, antiquarks and gluons) scatter 
and propagate in a self-generated scalar mean-field potential \cite{Cassing:2009vt}. 
On the partonic side the following elastic and inelastic
interactions are included $qq \leftrightarrow qq$, $\bar{q}
\bar{q} \leftrightarrow \bar{q}\bar{q}$, $gg \leftrightarrow gg$,
$gg \leftrightarrow g$, $q\bar{q} \leftrightarrow g$  exploiting
'detailed-balance' with temperature dependent cross sections
evaluated at the tree-level with the propagators and couplings from the DQPM.

The expansion of the system leads to a decrease of the local energy density and, once the local energy density becomes close to or lower than $\epsilon_c=0.5$ GeV/fm$^3$, the massive colored off-shell quarks and antiquarks hadronize to colorless off-shell mesons and baryons.
On the hadronic side, PHSD includes explicitly the baryon octet and
decouplet, the~$0^-$- and $1^-$-meson nonets as well as selected
higher resonances as in the Hadron--String--Dynamics (HSD)
approach~\cite{Cassing:1999es}. 
In the PHSD approach the full evolution of a relativistic heavy-ion collision, 
from the initial hard $NN$ collisions out of equilibrium up to the hadronisation 
and final interactions of the resulting hadronics, is described on the same footing. 
We recall that this approach has been successfully employed for $p+p$, $p+A$ and 
$A+A$ reactions from SIS to LHC energies
\cite{Cassing:2008sv,Cassing:2008nn,Cassing:2009vt,Bratkovskaya:2011wp,Linnyk:2015rco}.

\section{Strings in the PHSD}

In the PHSD/HSD  the string excitation and decay plays a decisive role 
for inelastic $BB$, $mB$, $mm$ collisions in a wide energy range. 
In the initial phase the high energy hadron-hadron
collisions are described by the Lund string model \cite{LUND}, 
where two incoming nucleons emerge the reaction as two excited color
singlet states, i.e. 'strings'. A string is characterized by the leading constituent
quarks of the incoming hadron as a string ends which are connected by a color flux 
tubs (color-electric field). The baryonic ($qq-q$) and mesonic ($q-\bar{q}$) strings are considered with different flavors ($q = u,d,s$). 
As the string ends recede, virtual $q\bar q$ or $qq\bar q\bar q$ pairs are produced in the
uniform color field by a tunnelling process (described by the Schwinger formula \cite{Schwinger}), causing the breaking of the string.
The produced quarks and antiquarks recombine with neighbouring partons 
to "prehadronic" states which will approach hadronic quantum states 
(mesons or baryon-antibaryon pairs) after a formation time $\tau_F\sim 0.8\,$fm/c 
(in the rest-frame of the string). 
$\tau_F$ is an internal PHSD parameter, introduced in Ref. \cite{Ehehalt:1996uq},  
for controlling the dynamics of "pre-hadronic" states in HIC, 
it is the same for all energies from SIS to LHC.
In the calculational frame of heavy-ion reactions (which is chosen to be the initial $NN$ center-of-mass frame) the formation time then is $t_F=\tau_F \cdot \gamma$, 
where $\gamma=1/\sqrt{1-v^2}$ and  $v$ is the velocity of the particle in the calculational frame.

The numerical realization of the Lund model in the PHSD is based
on the FORTRAN codes FRITIOF 7.02 \cite{FRITIOF}, which includes
PYTHIA 5.5, JETSET 7.3, ARIADNE 4.02,  for energies up to RHIC
and  PYTHIA 6.4 \cite{Sjostrand:2006za} with the Innsbruck pp tune (390) 
\cite{Innsbruck} with CTEQ5 LO PDFs (Jul 2013) for the LHC energies. 
A smooth transition between both descriptions is realized at  "intermediate" energies 
of $\sqrt{s_{NN}} \le 250$~GeV. 
In the PHSD the Lund programs are "tuned", i.e. adjusted in order to get a better agreement 
with experimental data for elementary collisions. 

It is important to stress here that there is {\it a conceptual difference} 
in the treatment of "free" $hh$ reactions (i.e. $hh$ collisions in the vacuum) in the PHSD 
and PYTHIA (or FRITIOF) beyond the "tuning" of Lund routines:
In the PHSD, contrary to PYTHIA, the elementary $hh$ collisions in the vacuum are simulated 
in a dynamical way, similar to $p+A$ or $A+A$ collisions, i.e. we follow the time evolution 
of $hh$ reactions -- starting from string excitations for high energy $hh$ reactions, 
to string fragmentations to hadrons, the propagation of hadrons and the dynamical
decay of baryonic and mesonic resonances during the expansion of the system.
Moreover, the produced hadrons can re-interact elastically and inelastically.
The inelastic reactions include the secondary less energetic $hh$ string 
excitations and low energy $hh$ collisions $2 \to n$ where $n=2,3,4...$
\cite{Cassing:1999es} as well as multi-meson fusion reactions to 
baryon-antibaryon pairs and backward reactions ($ n \ mesons \leftrightarrow B+\bar B$) \cite{Cassing:2001ds,Seifert:2017oyb}. 
The Lund routines (FRITIOF and PYTHIA) are used only as event generators for 
energetic inelastic collisions above a "string threshold" (defined below) 
which gives us the multiplicity and  momentum distribution of produced hadrons. 
The elastic scattering is realized according to the PHSD
routines. Also the decay of resonances - mesonic and baryonic - is realized by
the PHSD routines by playing Monte-Carlo for the decay probability with the
life-time which is inverse to the total width of the resonance.

Thus, for elementary $hh$ (i.e. $BB, mB, mm$) reactions in vacuum we solve microscopic 
transport equations for the propagation in time of all degrees of freedom 
with a collision term for their interactions. 
We note here that recently a framework for hadronic rescattering in $p+p$ collisions
has been proposed for PYTHIA in Ref. \cite{Sjostrand:2020gyg}.
The inclusion of final state interactions  can slightly change the final 
multiplicity of hadrons as compared to the production point by string decay, 
as well as their momentum distribution due to elastic scattering as will be
discussed in Section 4. 

We note that all discussed above is relevant for the PHQMD \cite{Aichelin:2019tnk}
approach, too, since the treatment of the collision integral in the PHQMD is identical 
to the PHSD. Technically speaking the PHQMD always merges with the latest
version of the PHSD and all development in modelling of collisions 
are incorporated in by the PHQMD automatically. Thus, in this study we will address 
the PHSD as a main laboratory for testing the string dynamics.

\subsection{"PHSD tune" of the string model}

Here we discuss the major changes of the Lund codes (FRITIOF 7.02 and PYTHIA 6.4)
for their application to the HICs study within the PHSD which we will call as
"the PHSD tune":\\
$\bullet$ 
We extend the applicability of string routines to lower energies by  
lowering the threshold from the default value of $\sqrt{s_{min}}=10$~GeV for the
minimal possible energy, 
to $\sqrt{s_{BB}}=2.65$~GeV for $BB$ collisions, $\sqrt{s_{mB}}=2.4$~GeV for $mB$ collisions
and $\sqrt{s_{mm}}=1.3$~GeV for $mm$ collisions. Even going much below the 
range of the 'default' model applicability, FRITIOF 7.02 and PYTHIA 6.4 give a very 
reasonable description of elementary collisions which we will demonstrate in the 
next section.

$\bullet$ 
At the string decay, the "flavour chemistry" of the produced quarks is determined
via the Schwinger mechanism \cite{Schwinger}, generalized to $q\bar q$ pairs 
in Refs. \cite{Casher:1978wy,Andersson:1980vj,Gurvich:1979nq},
which defines the production probability of massive $s\bar s$ pairs with
respect to light flavor  $(u\bar u,d\bar d)$ pairs:
\begin{equation}
\label{Schwinger-formula} 
\frac{P(s\bar s)}{P(u\bar u)}=\frac{P(s\bar s)}{P(d\bar
d)}=\gamma_s=\exp\Bigl(-\pi\frac{m_s^2-m_{u,d}^2}{2\kappa}\Bigr)\,,
\end{equation}
with $\kappa\approx 0.176\,$GeV$^2$ denoting the string tension while
$m_{u,d,s}$ are the constituent quark masses for strange
and light quarks. For the constituent quark masses
$m_u\approx0.35$ GeV and $m_s\approx0.5\,$GeV are adopted in the vacuum
which are selected in line with Dyson-Schwinger calculations 
\cite{Eichmann:2016yit}. 
From Eq. (\ref{Schwinger-formula}) follows that the production of strange quarks 
is thus suppressed by a factor of $\gamma_s\approx 0.3$ with respect to the light quarks,
which also is the default setting in the Lund routines.

While the strangeness production in proton-proton
collisions at SPS energies is reasonably well reproduced with this
value, the strangeness yield for $p~+~Be$ collisions at AGS energies
is underestimated by roughly 30\% (cf. \cite{Jgeiss}). For that
reason the relative factors used in the PHSD/HSD model are
\cite{Jgeiss}
\begin{eqnarray}
u:d:s:uu = \left\{
\begin{array}{ll}
1:1:0.3:0.07 &, \mbox{at SPS to RHIC } \\ 1:1:0.4:0.07 &,
\mbox{at AGS energies} ,
\end{array}
\right. \label{HSD-supp}
\end{eqnarray}
with a linear transition of the strangeness suppression factor
$\gamma_s$ as a function of $\sqrt{s_{NN}}$ in between. These settings
have been fixed in Ref. \cite{Jgeiss} for HSD in 1998 and kept since
then also for PHSD.

$\bullet$
We modify the flavour decomposition for the production of some mesonic states
-- $\eta, \eta^\prime, \rho, \omega, \phi$ and baryonic states $\bar p, 
\Delta^{++}$  in order to achieve a better agreement with available 
experimental data in $pp$ collisions. 
For that we changed the corresponding control parameters 
in the Lund routines and/or adjusted the hadronic final state directly, e.g. by 
letting some fraction of produced  vector mesons $\rho, \omega$ decay to pions
($\phi$ decay to kaons), i.e. taking the pions (kaons) as a final string decay products 
-- cf. PHSD Refs. \cite{Cassing:1999es,Bratkovskaya:2013vx,Bratkovskaya:2007jk}.

$\bullet$
The production of all charm and beauty states is realized according to the PHSD
routines and not adopted from PYTHIA - cf. \cite{Song:2015sfa}.

$\bullet$
The production of electromagnetic probes, direct photons and lepton pairs, 
is treated according to the PHSD routines - cf. \cite{Cassing:1999es,Bratkovskaya:2007jk}.

$\bullet$
A distribution of the newly produced hadrons in momentum space, i.e. the fraction 
of energy and momentum that they acquire from the decaying string, is defined by a fragmentation function $f(x,m_T)$. It gives the probability distribution for a hadron with transverse mass $m_T$ to be produced with an energy-momentum fraction $x$ from the fragmenting string:
\begin{eqnarray}
f(x,m_T)\approx {1 \over x} (1-x)^a \exp\left(-b m_T^2/x  \right),
\label{f_x}
\end{eqnarray}
where $a, b$ are parameters.
In the PHSD we use  $a=0.23$ and $b=0.34$~GeV$^{-2}$ \cite{Jgeiss}.
These settings for the string decay to hadrons have been found to match well experimental
observations for particle production in $p+p$ and $p+A$ reactions \cite{Cassing:1999es}. 

$\bullet$
In the standard version of FRITIOF/PYTHIA the baryonic and mesonic resonances 
are produced according to the non-relativistic Breit-Wigner spectral function with
a constant width. Moreover, the Breit-Wigner shape is truncated
symmetrically around the pole mass, $|M-M_0|<\delta$, with
$\delta$  chosen 'properly' for each particle such that no
problems are encountered in the particle decay chains.
In PHSD strings we incorporate  the fully relativistic Breit-Wigner spectral functions 
with mass dependent widths \cite{Bratkovskaya:2007jk}.
Also the truncation of the spectral function in mass is removed, 
i.e. the resonance mass is chosen within the
physical thresholds. As before the total energy and momentum conservation holds 
strictly in the extended Lund routines.

\subsection {‘In-medium’ extension of the Lund string model in the PHSD}

PHSD incorporates in-medium effects in the Lund string model, i.e. changes
of hadronic properties in a dense and hot environment as created in HICs.
The propagation of off-shell hadrons is realized by
the Cassing-Juchem off-shell transport equations based on the Kadanoff-Baym 
equations (cf. the review \cite{Cassing:2008nn}).

We note that the 'in-medium' modifications are {\it not used} in our present analysis 
for $pp$ collisions since the density of baryonic matter is rather low there, 
however, they become relevant for HIC's where string production happens 
in a hot and dense environment. Our discussion of 'in-medium' 
modifications here primarily addresses  readers interested in HIC results within the PHSD.

$\bullet$ We incorporate the in-medium spectral functions for mesonic 
and baryonic resonances in the Lund model by including the density dependent 
self-energy and in-medium width (depending on the local baryon density and temperature).
It allows to study in-medium effects such as collisional broadening of spectral 
functions of vector mesons ($\rho, \omega, \phi, a_1$)
\cite{Bratkovskaya:2007jk,Linnyk:2015rco}, which is mandatory for the description
of dilepton data from HICs.
Also it allows to study in-medium effects for the strange mesons 
$K, \bar K$ \cite{HSDK} and strange vector mesons $K^*, \bar K^*$ 
\cite{Ilner:2016xqr,Ilner:2017tab}. 

$\bullet$ The chiral symmetry restoration effect (CSR) has been incorporated in the 
PHSD  via the Schwinger mechanism for the string decay in the dense medium which
is formed by the primary collisions of nucleons and building of strings 
during the penetration  of the colliding nuclei.  In this initial phase a partial 
restoration of chiral symmetry occurs  which leads to a dropping of 
the scalar quark condensate in the hadronic environment of  finite baryon 
and meson density which can be estimated within the non-linear $\sigma-\omega$ model.
The dropping of the scalar quark condensate leads to a modification of the constituent 
quark masses for light and strange quarks and thus affect the "chemistry" of 
decaying strings via the Schwinger mechanism - cf. \cite{PHSD_CSR,Alessia}. 
This leads to an enhancement of stran\-geness production in the dense
baryonic medium before the deconfined phase may occur.

$\bullet$ We take into account the initial state Cronin effect which we
model in a dynamical way, i.e. $<k_T^2>$ the average transverse momentum squared 
of the partons in the nuclear medium created in $p+A$ or $A+A$ collisions, 
is enhanced due to induced initial semi-hard gluon radiation in the medium, 
which is not present in the vacuum due to the constraint of color neutrality 
\cite{Cassing:2003sb}.


\section{Comparison of PHSD and PYTHIA 8.2 results for $p+p, p+n, n+n$ reactions}

In this section we present a comparison of the results from the PHSD approach 
within the "PHSD tune" of strings to the default PYTHIA 8.2 (in 'Soft QCD' mode)
\cite{PYTHIA82} 
for elementary $p+p, p+n$ and $n+n$ reactions.
We also compare both models to the experimental data when available.
We note that in spite that the most experimental data exist for $p+p$ collisions 
only, it is very important to have reliable results for other isospin 
channels as $p+n$ and $n+n$ since such reactions are more frequent in HICs 
due to the larger number of neutrons compared to protons in heavy nuclei.
We note that all PHSD results shown here are computed including final state 
hadronic rescattering (FSI), except of special examples which we will 
discuss below.

\subsection{Hadron Multiplicities vs $\sqrt{s_{NN}}$}

      \begin{figure*}[!ht]
        \centering
        \resizebox{0.8\textwidth}{!}{
          \begin{tabular}{cc}
            \includegraphics[page=1]{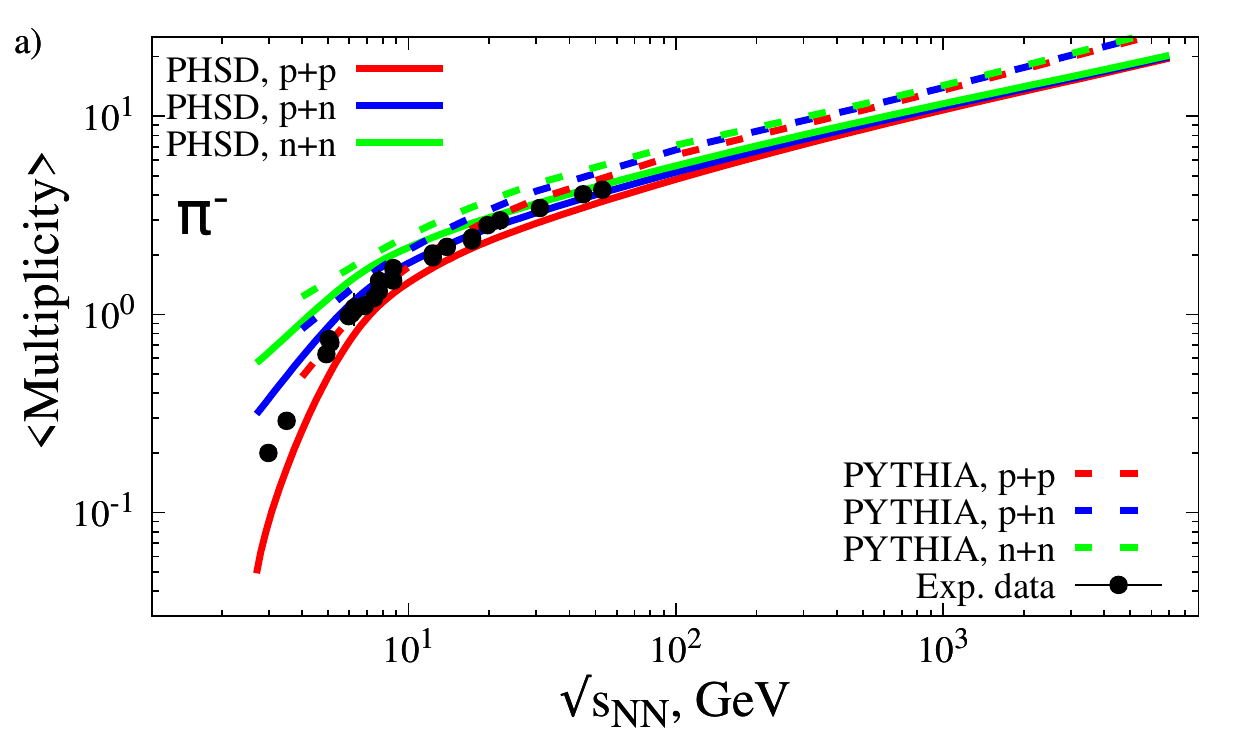} &
            \includegraphics[page=2]{phqmd_pythia_FSI.pdf} \\
            \includegraphics[page=3]{phqmd_pythia_FSI.pdf} &
            \includegraphics[page=4]{phqmd_pythia_FSI.pdf} \\
            \includegraphics[page=7]{phqmd_pythia_FSI.pdf} &
            \includegraphics[page=6]{phqmd_pythia_FSI.pdf} \\
            \includegraphics[page=5]{phqmd_pythia_FSI.pdf} &
            \includegraphics[page=8]{phqmd_pythia_FSI.pdf} \\
          \end{tabular}
        }
\caption{Total multiplicities of $\pi^{\pm}$,  $K^{\pm}$, $p$ , $\bar{p}$, 
$K^{0}_{S}$ and $\Lambda+\Sigma^{0}$  produced in $N+N$ collisions: 
red lines corresponds to $p+p$, blue lines to $p+n$ and green lines to $n+n$ reactions.
The PHSD results are drawn by solid lines, the PYTHIA 8.2 results by dashed lines.
The black dots indicate the experimental data for $p+p$ collisions from Refs. \cite{na61_hminus}--\cite{antinucci:1973}.}
        \label{mult_pn}
      \end{figure*}
      \begin{figure*}[!htbp]
        \centering
        \resizebox{0.8\textwidth}{!}{
          \begin{tabular}{cc}
            \includegraphics[page=1]{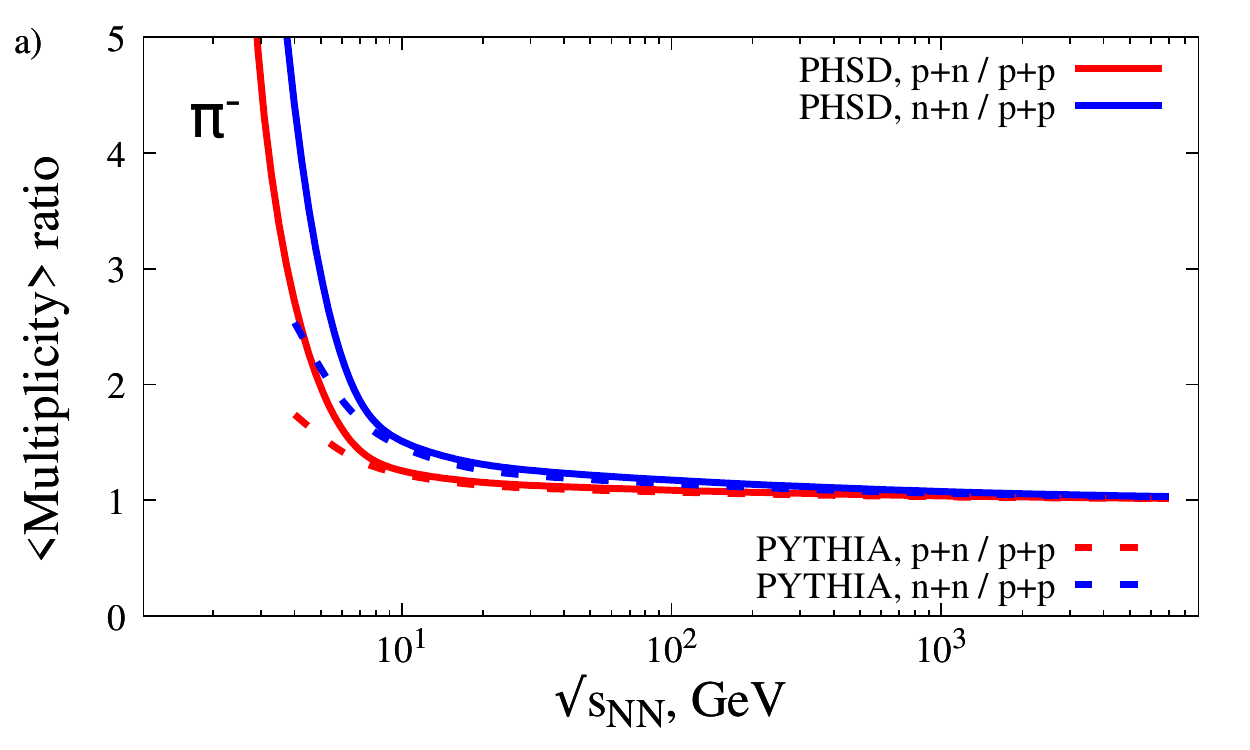} &
            \includegraphics[page=2]{phqmd_ratio_linear_FSI.pdf} \\
            \includegraphics[page=3]{phqmd_ratio_linear_FSI.pdf} &
            \includegraphics[page=4]{phqmd_ratio_linear_FSI.pdf} \\
            \includegraphics[page=7]{phqmd_ratio_linear_FSI.pdf} &
            \includegraphics[page=6]{phqmd_ratio_linear_FSI.pdf} \\
            \includegraphics[page=5]{phqmd_ratio_linear_FSI.pdf} &
            \includegraphics[page=8]{phqmd_ratio_linear_FSI.pdf} \\
          \end{tabular}
        }
        \caption{Ratio of $\pi^{\pm}$ $K^{\pm}$, $p$ , $\bar{p}$, $K^{0}_{S}$ 
        and $\Lambda+\Sigma^{0}$ multiplicities in $"p+n" / "p+p"$ reactions 
        (red lines) and  in $"n+n" / "p+p"$ (blue lines) calculated within the PHSD
        (solid lines) and PYTHIA 8.2 (dashed lines).}
        \label{sys_ratio}
      \end{figure*}
We start with a comparison of the excitation function of the total multiplicities 
(i.e. $"4\pi"$ - without any cuts on rapidity etc.) of 
$\pi^\pm, K^\pm, p, \bar p, K^0_S, \Lambda+\Sigma^0$
in $p+p$, $p+n$ and $n+n$ collisions as presented in Fig. \ref{mult_pn}.
The PHSD calculations cover  the energy range $\sqrt{s_{NN}}=2.7 - 7000$~GeV 
and PYTHIA 8.2 -- $\sqrt{s_{NN}}=4 - 7000$ GeV. (We lower the default PYTHIA 8.2
threshold of 10 GeV in view of a closer comparison with the PHSD results.)
The red lines correspond to $p+p$ collisions, blue lines to $p+n$ 
reactions and green lines  to $n+n$. The PHSD results are drawn with solid lines, 
PYTHIA 8.2 results with dashed lines.
The black dots represent the experimental data  for $p+p$ collisions 
\cite{na61_hminus}--\cite{antinucci:1973}.

One can see that i) PYTHIA 8.2 provides systematically larger multiplicities
for pions, protons and especially  $\bar p$. 
ii) Furthermore, one can see the rather strong isospin dependence of hadron 
multiplicities in $p+p$, $p+n$, $n+n$ reactions: the multiplicities of hadrons
in $p+p$ reactions are larger than in $p+n$ and $n+n$ reactions.
This is demonstrated in Fig. \ref{sys_ratio} which shows the ratios 
of $\pi^{\pm}$,  $K^{\pm}$, $p$ , $\bar{p}$, 
$K^{0}_{S}$ and $\Lambda+\Sigma^{0}$ multiplicities in different reactions: 
the red lines indicate the $"p+n" / "p+p"$ ratio, while the blue lines the 
$"n+n" / "p+p"$ ratio. Here again the solid lines show the PHSD calculations  while 
the dashed lines indicate the PYTHIA results.
One can see that both models give very similar ratios which indicate the same flavour decomposition according to isospin channels. 
The ratios of produced hadrons approach to 1 with increasing energy, i.e.
at $\sqrt{s_{NN}}\ge 10\div 30$ GeV. However, at low energies there is a strong 
isopin  dependence due to the initial combination of charges and flavours.

\begin{figure*}[!htbp]
        \centering
        \resizebox{0.8\textwidth}{!}{
          \begin{tabular}{cc}
            \includegraphics[page=1]{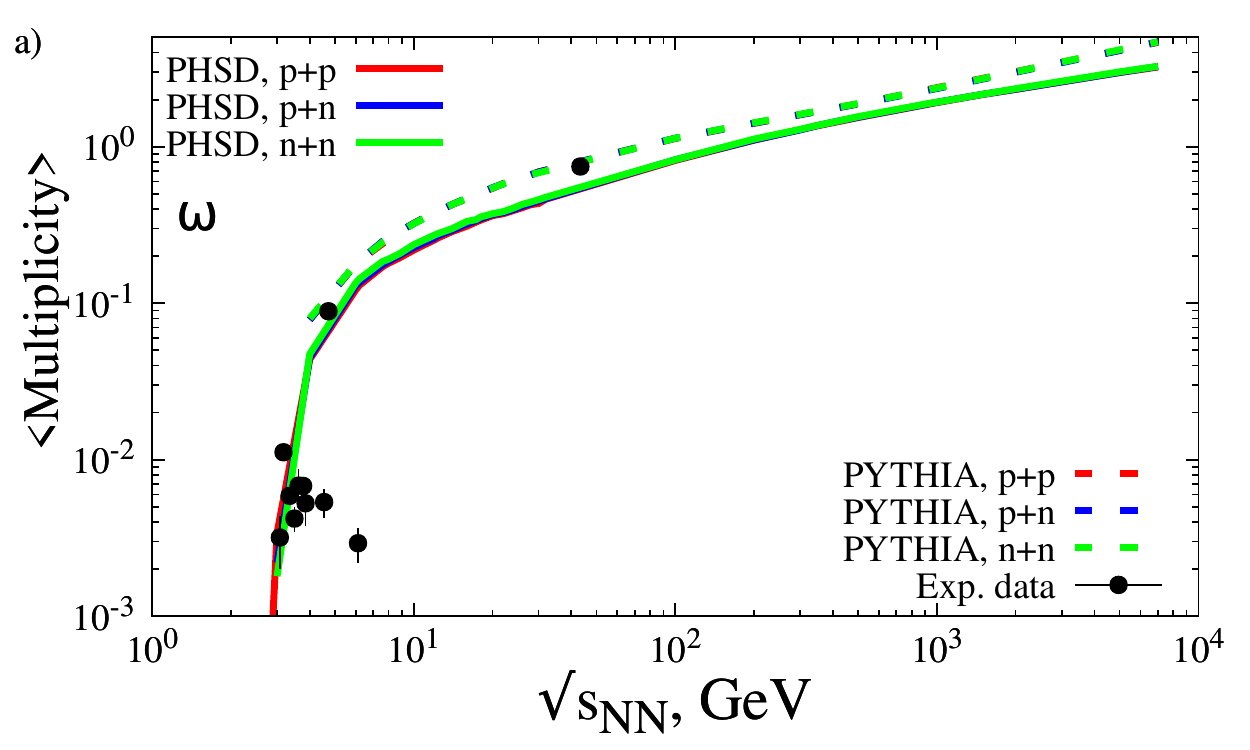} &
            \includegraphics[page=2]{phqmd_pythia_vecmes.pdf} \\
            \includegraphics[page=3]{phqmd_pythia_vecmes.pdf} &
            \includegraphics[page=4]{phqmd_pythia_vecmes.pdf} \\
            \includegraphics[page=5]{phqmd_pythia_vecmes.pdf} &
            \includegraphics[page=6]{phqmd_pythia_vecmes.pdf} \\
            \includegraphics[page=9]{phqmd_pythia_vecmes.pdf} &
            \includegraphics[page=8]{phqmd_pythia_vecmes.pdf} \\
          \end{tabular}
        }
\caption{Total multiplicities of vector mesons $\omega$ , $\rho^{\pm}$, $\rho^{0}$ 
$K^{*\pm}$ and $\phi$ produced in inelastic collisions. 
The red lines correspond to $p+p$, blue lines to $p+n$ and green lines  
to $n+n$ collisions.
The PHSD results are drawn by solid lines, PYTHIA 8.2 results with dashed lines.
The black dots show of the experimental data for $p+p$ collisions ~\cite{Landolt:1988}.}
        \label{mult_vecmes}
      \end{figure*}

We continue with Fig. \ref{mult_vecmes} where we show the total ($"4\pi"$) multiplicity of 
vector mesons $\omega$, $\rho^{\pm}$, $\rho^{0}$, $\phi$ and strange vector mesons
$K^{*\pm}$, $K^{*0}$ produced in $N+N$ collisions: 
red lines corresponds to $p+p$, blue lines to $p+n$ and green lines  to $n+n$ reactions.
The PHSD results are drawn by solid lines, the PYTHIA 8.2 results by dashed lines.
The black dots represents the experimental data for $p+p$ collisions from 
Ref. \cite{Landolt:1988}. The isospin dependence is rather weak here in both models.
The multiplicities of light vector mesons are lower in PHSD  since they were corrected
for  better matching of existing data. This is also cross-checked by dilepton data
for $p+p$ as well as for HICs since the direct decay of vector mesons is one 
of the dominant channels for dilepton production for invariant masses $0.4\le M\le 1.2$ GeV
\cite{Bratkovskaya:2013vx,Linnyk:2015rco}.
On the contrary, the multiplicity of strange vector mesons in PHSD is larger
at high energies; the description of existing experimental data on $K^{0*}$ 
total multiplicities at intermediate energies \cite{Landolt:1988} is worth 
in PHSD than in PYTHIA 8.2.
On the other hand, the PHSD $p_T$ spectra of $K^*$ 
from $p+p$ at midrapidity at RHIC and LHC energies are in a good agreement 
with experimental data \cite{Ilner:2016xqr,Ilner:2017tab}. 

  
\begin{figure*}[!htbp]
        \centering
         \resizebox{0.7\textwidth}{!}{
          \begin{tabular}{cc}
            \includegraphics[page=1]{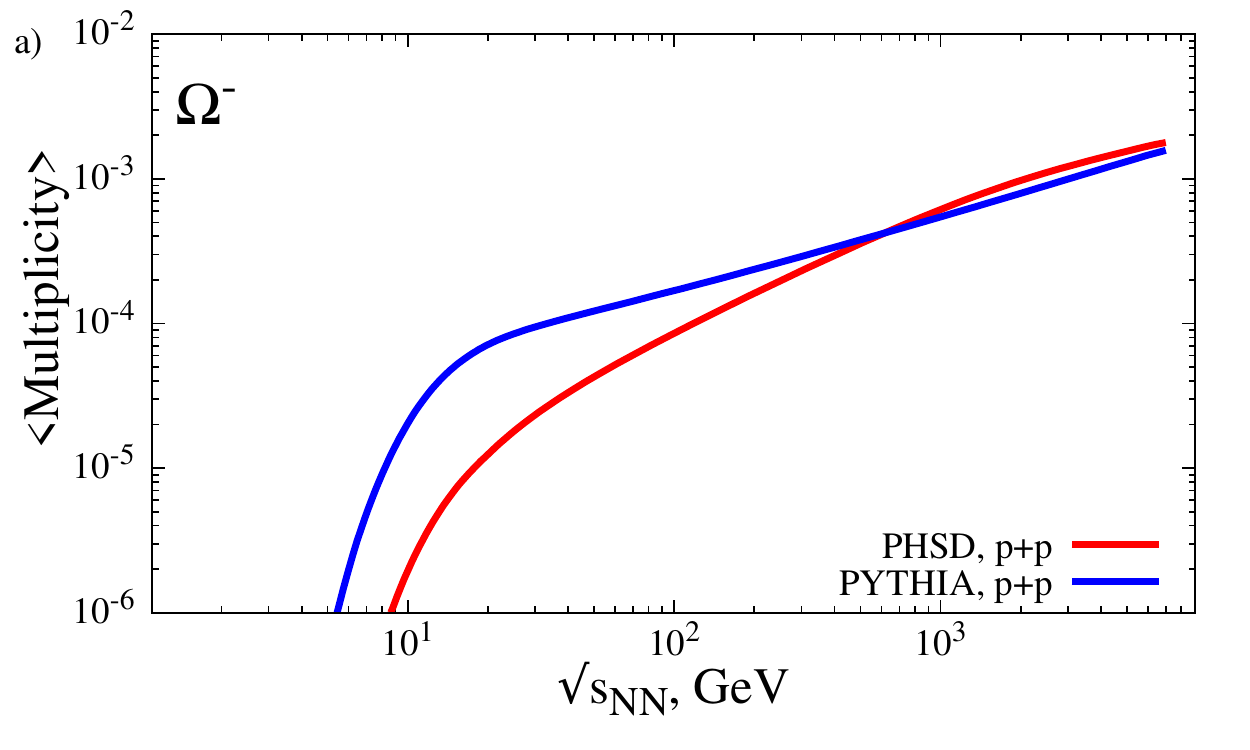} &
            \includegraphics[page=2]{dblstr_FSI.pdf} \\
            \includegraphics[page=3]{dblstr_FSI.pdf} &
            \includegraphics[page=4]{dblstr_FSI.pdf} \\
          \end{tabular}
        }
\caption{Total multiplicities of multi-strange baryons 
$\Omega^{-}, \bar\Omega^{+}, \Xi^{-}, \bar\Xi^{+}$ produced in $p+p$ collisions.
The red lines stand for the PHSD calculations while the blue lines shows PYTHIA 8.2 results. 
The black dots show the NA61/SHINE data \cite{Aduszkiewicz:2020dyw}.}
        \label{mult_dblstr}
      \end{figure*}

Figure \ref{mult_dblstr} shows the excitation function of the total multiplicity of 
multi-strange baryons $\Omega^{-}, \bar\Omega^{+}, \Xi^{-}, \bar\Xi^{+}$  
produced in $p+p$ collisions. The red lines stand for the PHSD calculations 
while the blue lines show PYTHIA 8.2 results. The deviations between both models
are rather large at low energies especially for $\Omega$ baryons.
More $"4\pi"$ experimental data are needed to construct the multi-strange baryon production.

 
  \subsection{Hadronic final state interactions (FSI) in $N+N$ reactions within the PHSD }
  
      \begin{figure*}[!htbp]
        \centering
         \resizebox{0.7\textwidth}{!}{
          \begin{tabular}{cc}
            \includegraphics[page=1]{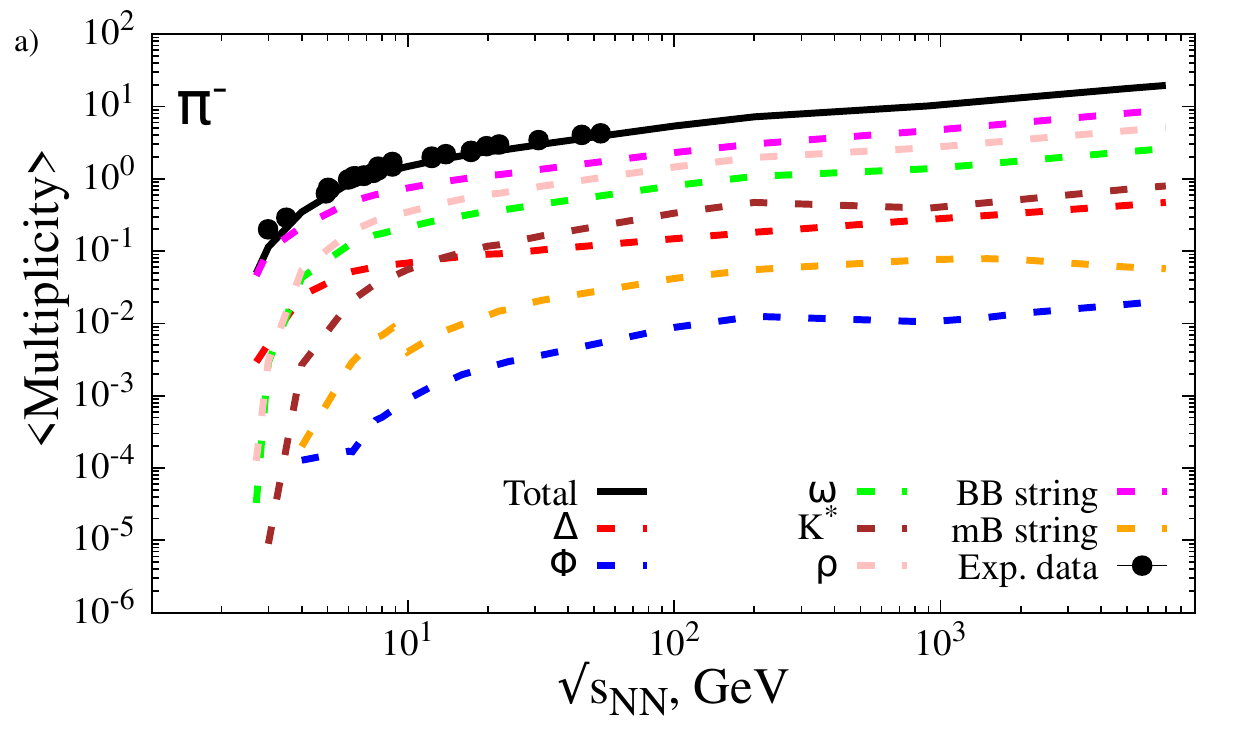} &
            \includegraphics[page=2]{phqmd_channels_FSI.pdf} \\
            \includegraphics[page=3]{phqmd_channels_FSI.pdf} &
            \includegraphics[page=4]{phqmd_channels_FSI.pdf} \\
          \end{tabular}
        }
\caption{Channel decomposition for  $\pi^{\pm}$ and $K^{\pm}$ production in $p+p$ collisions: 
The magenta lines ('$BB$ string') show the contribution to the total multiplicity 
from the direct hadron production from $BB$ string fragmentation, the orange lines  
from secondary $mB$ strings, while the lines $\Delta, \omega, K^*, \rho, \phi$ 
indicate the contribution from the decay of the corresponding resonances.   
The black dots indicate the experimental data for $p+p$ collisions from Refs. \cite{na61_hminus}--\cite{antinucci:1973}.    
        }
        \label{mult_mes}
      \end{figure*}
In order to demonstrate the production mechanisms of the stable final hadrons 
in PHSD  we present in Figure~\ref{mult_mes} the channel decomposition for  
$\pi^{\pm}$ and $K^{\pm}$ production in $p+p$ collisions: 
The magenta lines ('$BB$ string') show the contribution to the total multiplicity 
from the direct hadron production from $BB$ string fragmentation, formed by baryon-baryon collisions ($B=p, n, \Delta, ...$), the orange lines  
from secondary $mB$ strings, formed by meson-baryon collisions, 
while the lines '$\Delta, \omega, K^*, \rho, \phi$ '
indicate the contribution from the decay of corresponding resonances.
One can see that only about half of the final mesons come directly from 
$BB$ string fragmentation while the other half comes from resonance decays
and even secondary production channels (as $mB$ string, indicated here). Moreover,
the produced particles can scatter elastically or participate in charge exchange reactions.
Thus, in view of the final state hadronic interactions the dynamics of $N+N$ collisions 
are rather similar to the dynamics of HICs - the hadrons are produced at different times 
from different sources and not from a single vertex of the initial $N+N$ collision. 
Indeed, the total multiplicities in $N+N$ is much lower than in HIC at the same energies,
i.e. the density of particles is much smaller, correspondingly, the role
of final state interactions is strongly reduced. 
 
\begin{figure*}[!htbp]
        \centering
        \resizebox{0.7\textwidth}{!}{
          \begin{tabular}{cc}
            \includegraphics[page=1]{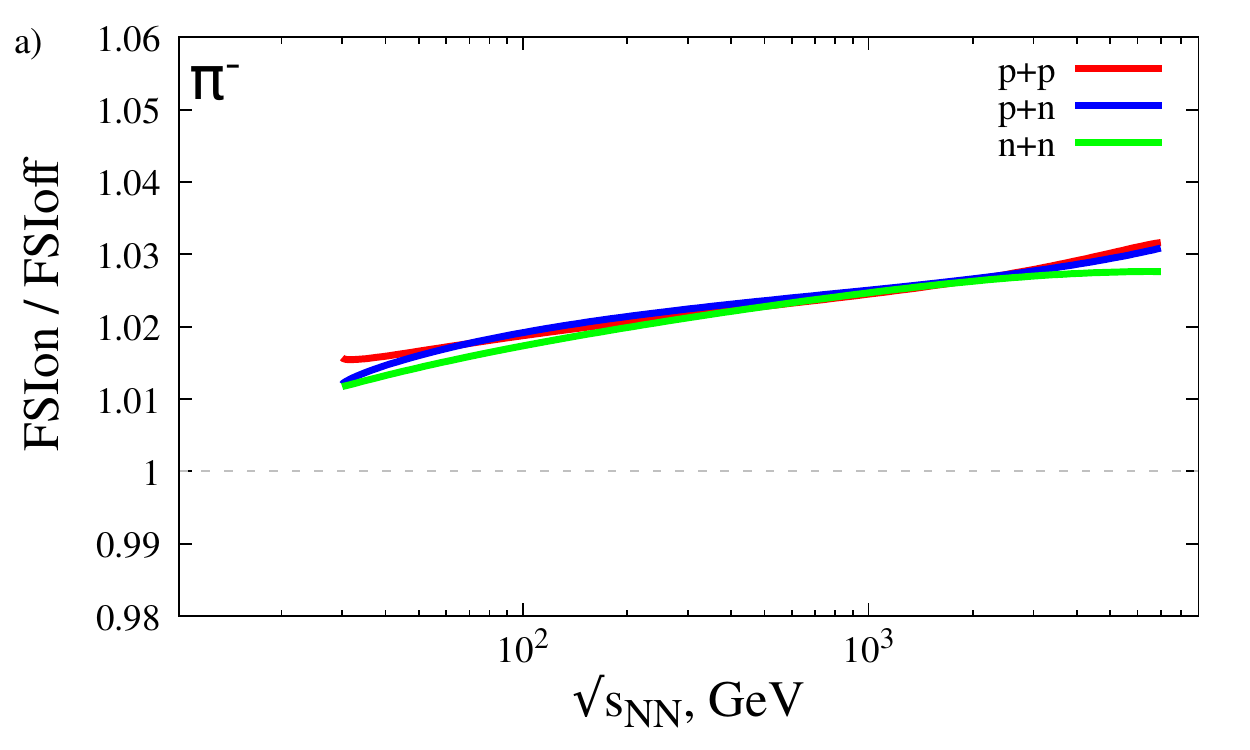} &
            \includegraphics[page=2]{ratios_FSI.pdf} \\
            \includegraphics[page=3]{ratios_FSI.pdf} &
            \includegraphics[page=4]{ratios_FSI.pdf} \\
            \includegraphics[page=7]{ratios_FSI.pdf} &
            \includegraphics[page=6]{ratios_FSI.pdf} \\
            \includegraphics[page=5]{ratios_FSI.pdf} &
            \includegraphics[page=8]{ratios_FSI.pdf} \\
          \end{tabular}
        }
        \caption{Ratios of total multiplicities with FSI ('FSIon') and
without FSI ('FSIoff') of $\pi^{\pm}$, $K^{\pm}$, $p$ , $\bar{p}$, 
$K^{0}_{s}$ and $\Lambda+\Sigma^{0}$  produced in $N+N$ collisions: 
the red lines correspond to $p+p$, blue lines -- to $p+n$ and green lines -- to $n+n$ reactions.}
        \label{ratio_fsi}
      \end{figure*}

In order to quantify the role of FSI in elementary $N+N$ reactions we perform  
PHSD calculations without FSI ('FSIoff') and compute the ratio of total multiplicities
with FSI ('FSIon') and without FSI. The results for the ratio 'FSIon/FSIoff'
are presented in Fig. \ref{ratio_fsi} for $\pi^{\pm}$, $K^{\pm}$, $p$ , $\bar{p}$, 
$K^{0}_{s}$ and $\Lambda+\Sigma^{0}$ produced in $N+N$ collisions: 
the red lines correspond to $p+p$, blue lines to $p+n$ and green lines to $n+n$ reactions.  One can see that with increasing energy the role of FSI increases
and reaches a few percent ($<5$\%) at the LHC energies. Moreover, the ratios show only 
a very small dependence on isospin channels $p+p$, $p+n$ or $n+n$.

\subsection{$x_{F}$ distributions at $\sqrt{s_{NN}}=17.3$ GeV}  

      \begin{figure*}[!htbp]
      \centering
      \resizebox{1.02\textwidth}{!}{
        \begin{tabular}{lll}
         \includegraphics[scale=1.2]{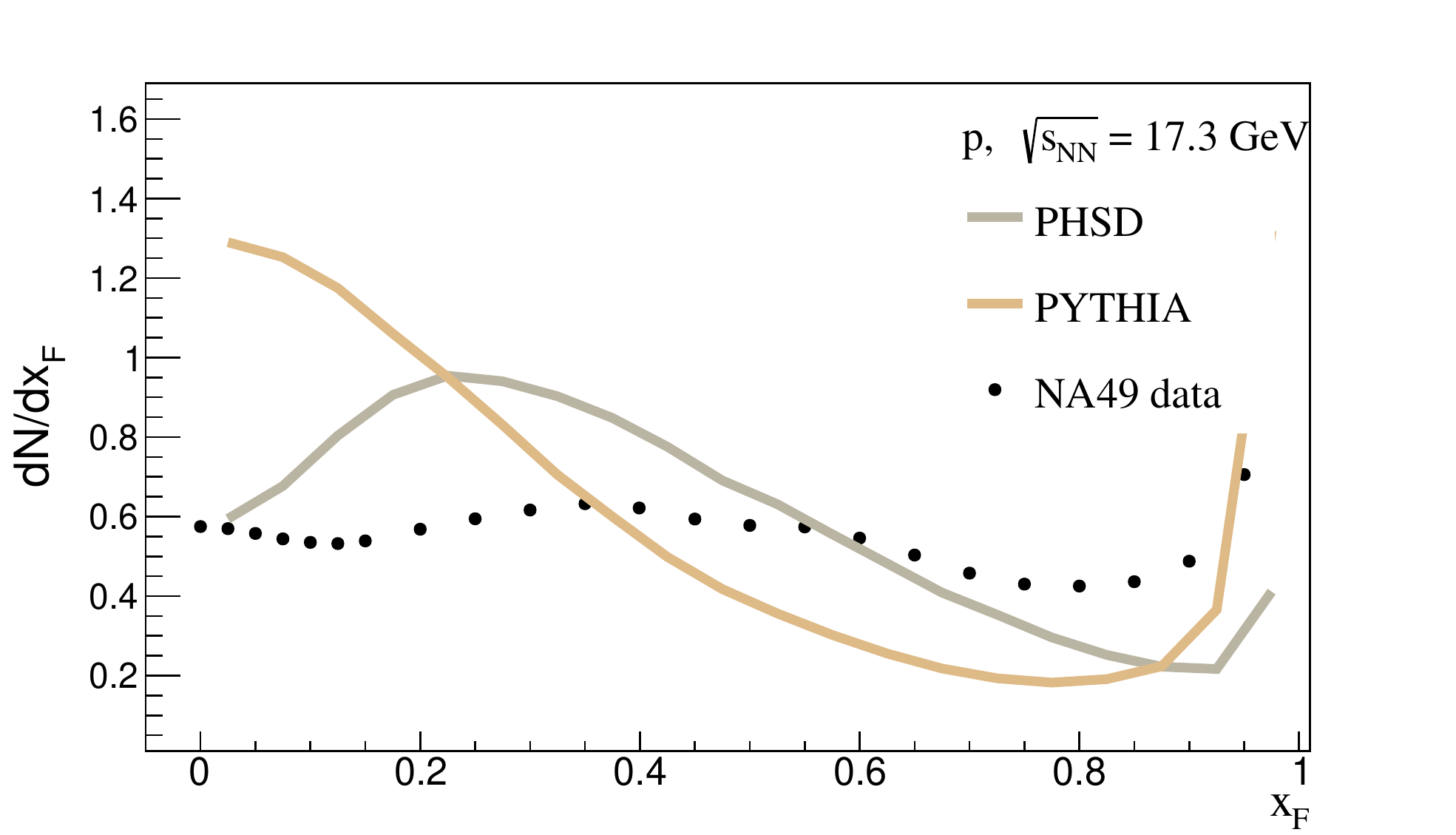} &
          \includegraphics{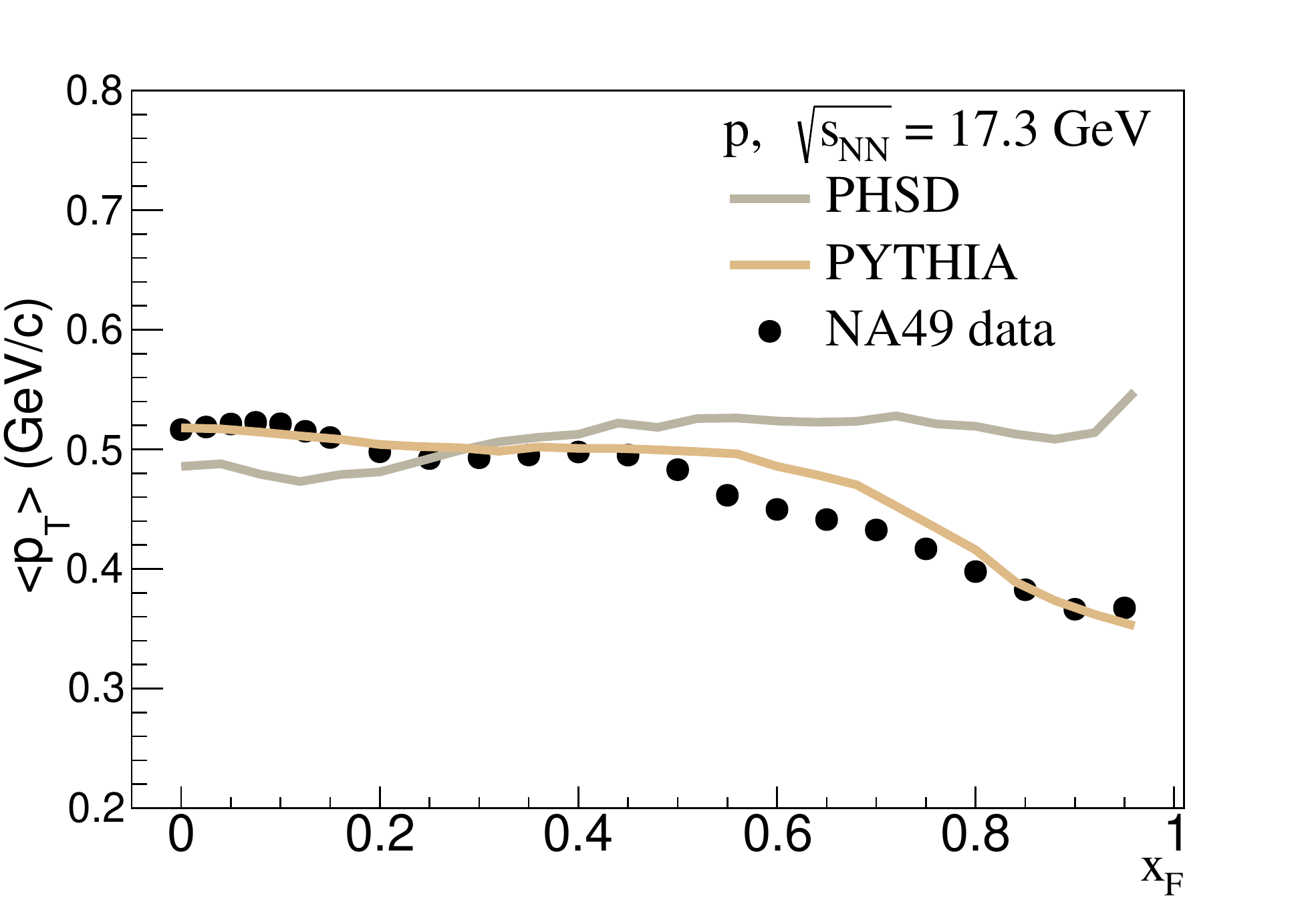} &
          \includegraphics{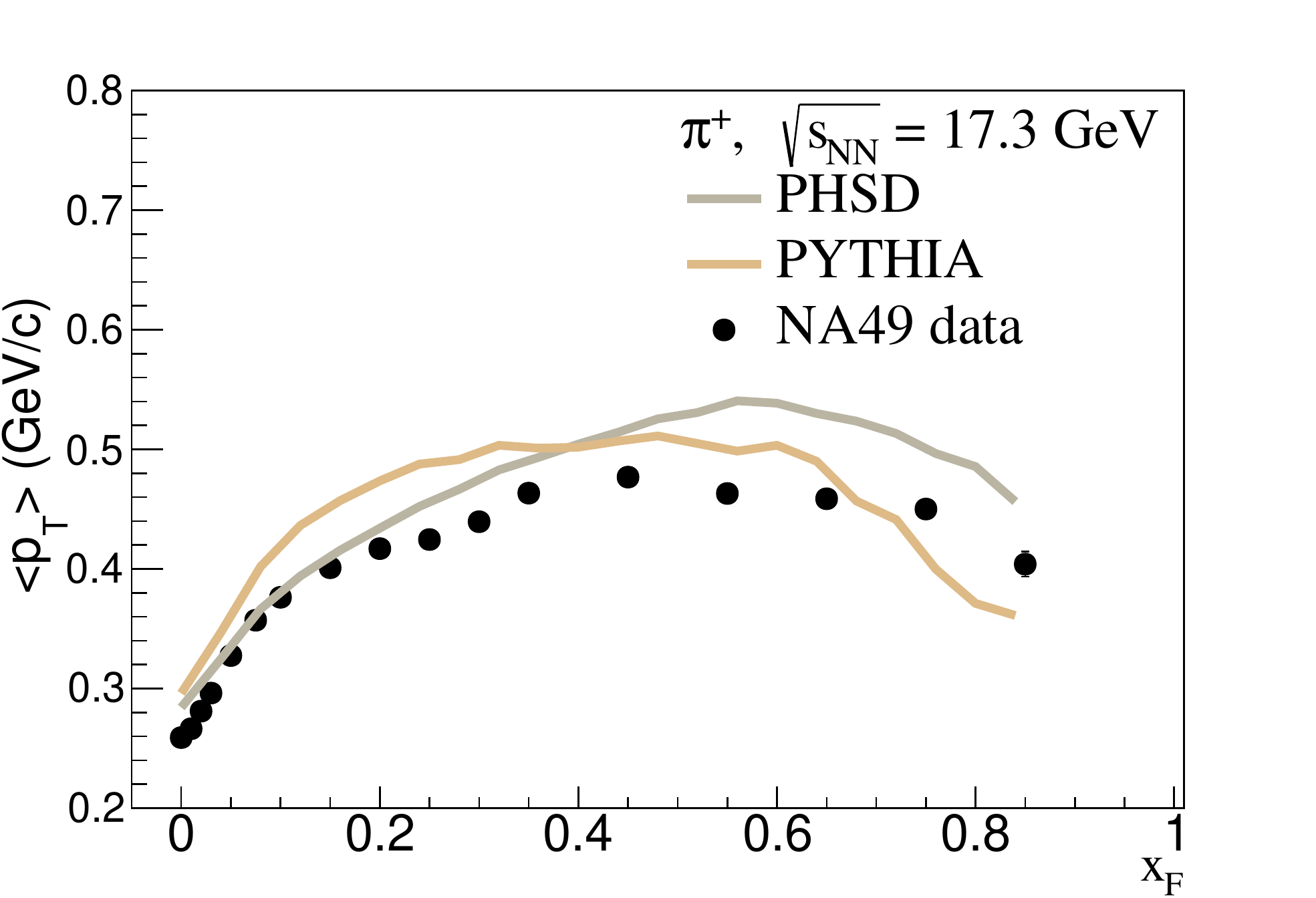} \\
        \end{tabular}
      }
      \caption{
Proton $x_{F}$ distribution  (left plot) in $p + p$ collisions at $\sqrt{s_{NN}}=17.3$ GeV. 
 Mean  transverse  momentum $<p_T>$ of protons (middle plot) and $\pi^{+}$ (right plot)
  as a function of $x_{F}$ in $p + p$ collisions at $\sqrt{s_{NN}}=17.3$ GeV. 
 The experimental data are taken from the NA49 Collaboration \cite{na49_ppbar,na49_pions}.}
      \label{xf_mpt}
    \end{figure*}
Now we step on to a comparison of differential observables at different energies.

We start with the comparison of the PYTHIA 8.2 results (orange lines) with the PHSD
results (grey lines) and the NA49 data~\cite{na49_ppbar,na49_pions} 
on proton $x_{F}$ distributions (left plot),  averaged transverse momentum $<p_T>$  of protons (middle plot) and $\pi^{+}$ (right plot) as a function 
of $x_{F}$ in $p + p$ collisions at $\sqrt{s_{NN}}=17.3$ GeV. 
One can see that the $dN/dx_F$ spectra are not well reproduced by both models,
on the other hand, the $<p_T>$ of protons agrees better with PYTHIA while 
the shape of $<p_T>$ of pions is approximately reproduced by both models.

We mentioned that the shape of $dN/dx_F$ distribution of protons is sensitive to 
the form fragmentation function - Eq. (\ref{f_x}). It turns quite non-trivial to 
fit the parameters in fragmentation function  in a way that it describes 
the experimental data on $dN/dx_F$ distribution at $\sqrt{s_{NN}}=17.3$ GeV and  
simultaneously keep the good description on other observables ($y, p_T$- spectra)
at different energies with the same parameters.
This require further developments as from theoretical side as well as more
experimental information on differential  $dN/dx_F$ distribution at different energies.

\subsection{Comparison of rapidity distributions at $\sqrt{s_{NN}}=6.2-17.3$ GeV}

     \begin{figure*}[!htbp]
      \centering
      \resizebox{0.8\textwidth}{!}{
        \includegraphics[scale=1]{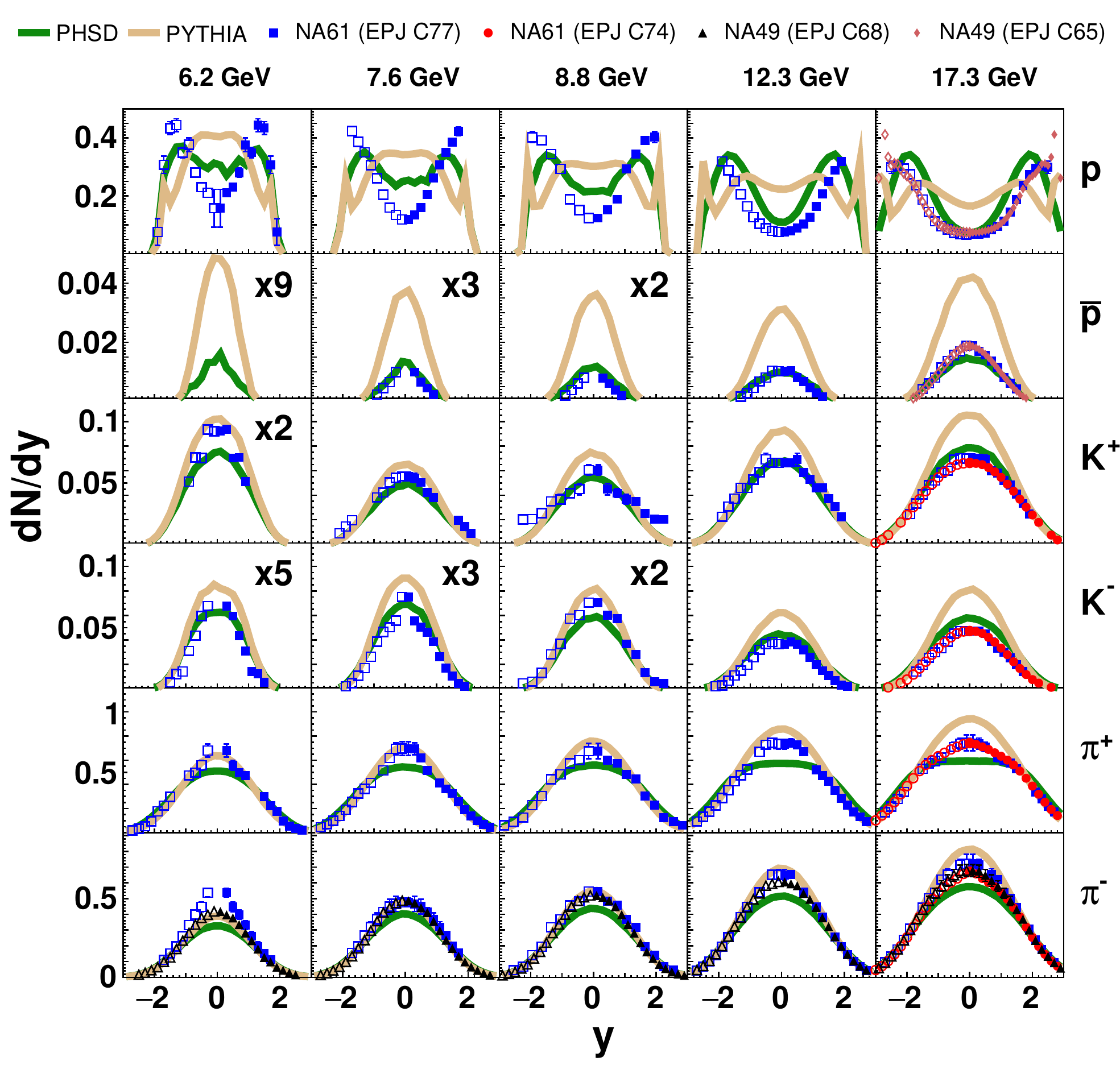}     }   
\caption{Rapidity distribution of protons, anti-protons, $K^\pm$ and 
$\pi^\pm$ from $p+p$ collisions at 6.2 GeV, 7.6 GeV, 8.8 GeV, 12.3 GeV, 17.3 GeV.
The PHSD results are presented by green solid lines while 
the PYTHIA 8.2 results are shown by brown solid lines.      
The experimental data  are taken from the NA61/SHINE~\cite{na61_hadrons}
and NA49~\cite{na49_ppbar} Collaborations.
The scaling factors  for the data and theoretical results 
are introduced for better visualization: 
for 6.2 GeV : $\bar p \times 9$, \ $K^+ \times 2$, \  $K^- \times 5$; \ 
for 7.6 GeV : $\bar p \times 3$,  $K^- \times 3$; \ 
and for 8.8 GeV: $\bar p \times 2$, \ $K^- \times 2$. \ 
}        
       \label{y_prot}               
    \end{figure*}

      \begin{figure}[!htbp]
        \centering
       \resizebox{0.52\textwidth}{!}{
            \includegraphics{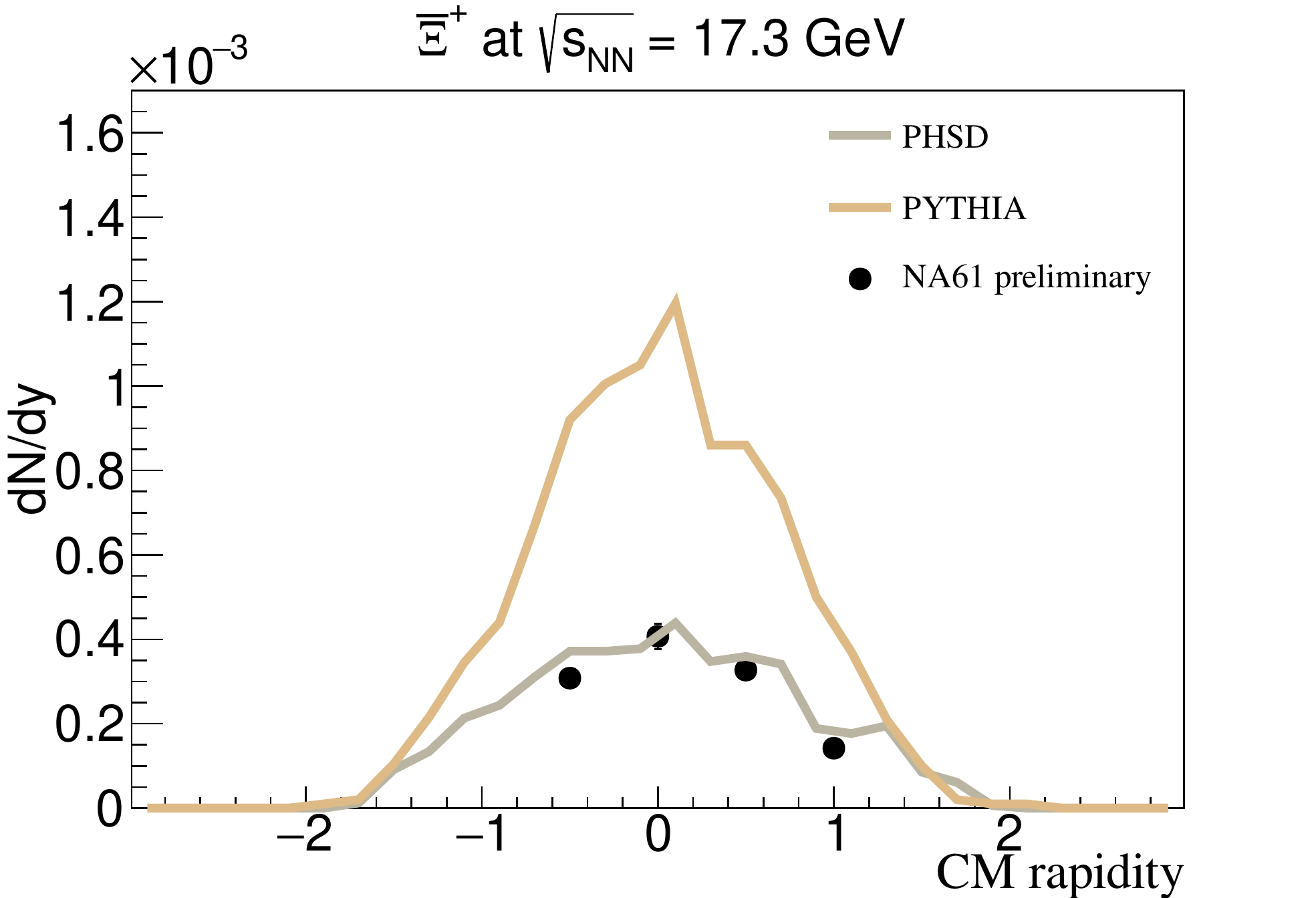} 
            \includegraphics{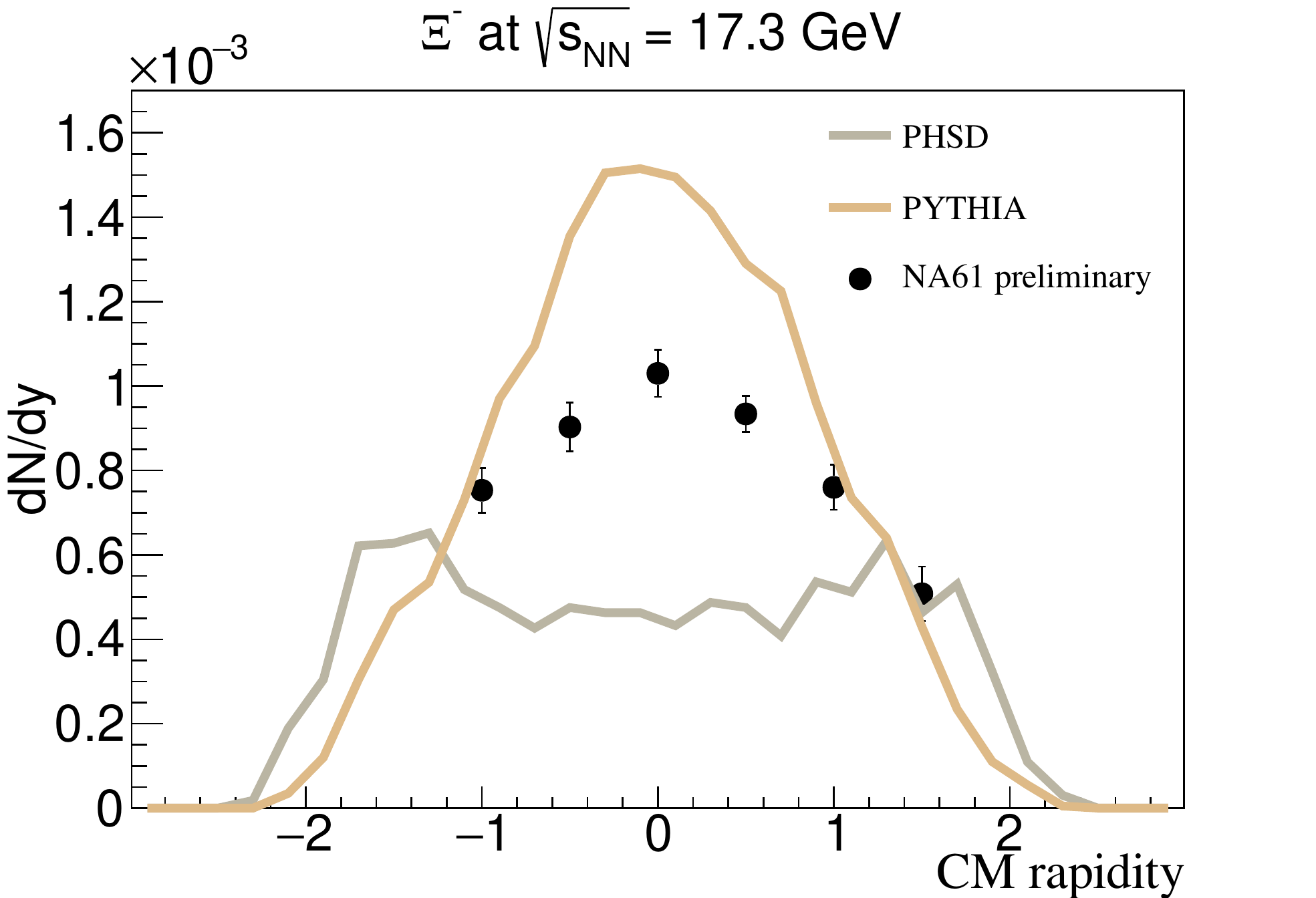} 
        }
\caption{Rapidity distribution of $\bar{\Xi}^{+}$ (left plot) and $\Xi^{-}$ (right plot) 
from $p+p$ collisions at 17.3 GeV.  
The PHSD results are presented by gray solid lines while 
the PYTHIA 8.2 results are shown by brown solid lines.
The experimental data  are taken from the NA61/SHINE Collaboration \cite{Aduszkiewicz:2020dyw}.}
        \label{dblstr_rapidity}
      \end{figure}
      
We continue with a comparison of the PHSD and PYTHIA 8.2 results to the experimental 
data from the NA49 and NA61/SHINE Collaborations on rapidity distributions  $dN/dy$
of protons, antiprotons, $\pi^\pm, K^\pm$ at $\sqrt{s_{NN}}$ = 6.2, 7.6, 8.8, 12.3, 17.3 GeV 
which  are shown in Fig.~\ref{y_prot}.
The data  from the NA61/SHINE \cite{na61_hminus,na61_hadrons} and 
NA49~\cite{na49_pions,na49_kaons,na49_ppbar} Collaborations are drawn by 
solid symbols, the open symbols indicate the data reflected about 
midrapidity. The PHSD results are plotted by solid lines. There are no 
experimental data for antiprotons below 8.8 GeV.
Additionally, Fig. \ref{dblstr_rapidity} shows the comparison
of model calculations for the rapidity distribution $dN/dy$ 
as a function of center-of-mass rapidity $y$ of $\bar{\Xi}^{+}$ (left plot) and $\Xi^{-}$ (right plot) from $p+p$ collisions at 17.3 GeV  to the experimental data 
from the NA61/SHINE Collaboration \cite{Aduszkiewicz:2020dyw}.

The model discrepancies with respect to the experimental data on the $dN/dy$ distributions
of newly produced hadrons can be attributed to a large extend to the description 
of proton "stopping", i.e.
to the shape of the rapidity distribution of protons. PYTHIA 8.2 tends to
have much stronger stopping at all considered energies, the proton
$dN/dy$ distributions are rather flat at midrapidity while the PHSD results show  
minima at midrapidity and a rise at target/projectile rapidity 
in line with the experimental data. Thus, the hadronic rapidity distributions
from PYTHIA 8.2 systematically overestimate the data while the PHSD results 
are closer to the data. However, this correlation is not so direct, e.g.
the PYTHIA results are perfectly on the data for $\pi^-$ at 
$\sqrt{s_{NN}}$ = 6.2, 7.6, 8.8, 12.3 GeV  while PHSD underestimates the data. 
The same holds for the description of multi-strange baryons $\bar{\Xi}^{+}$and $\Xi^{-}$
in Fig. \ref{dblstr_rapidity}. The latter  require further improvements on the 
mechanisms of multi-strangeness production at such intermediate energies.


\subsection{Comparison of transverse momentum $p_{T}$ spectra at $\sqrt{s_{NN}}=6.2 - 17.3$~GeV}

    \begin{figure*}[!htbp]
      \centering
      \resizebox{0.8\textwidth}{!}{
        \includegraphics{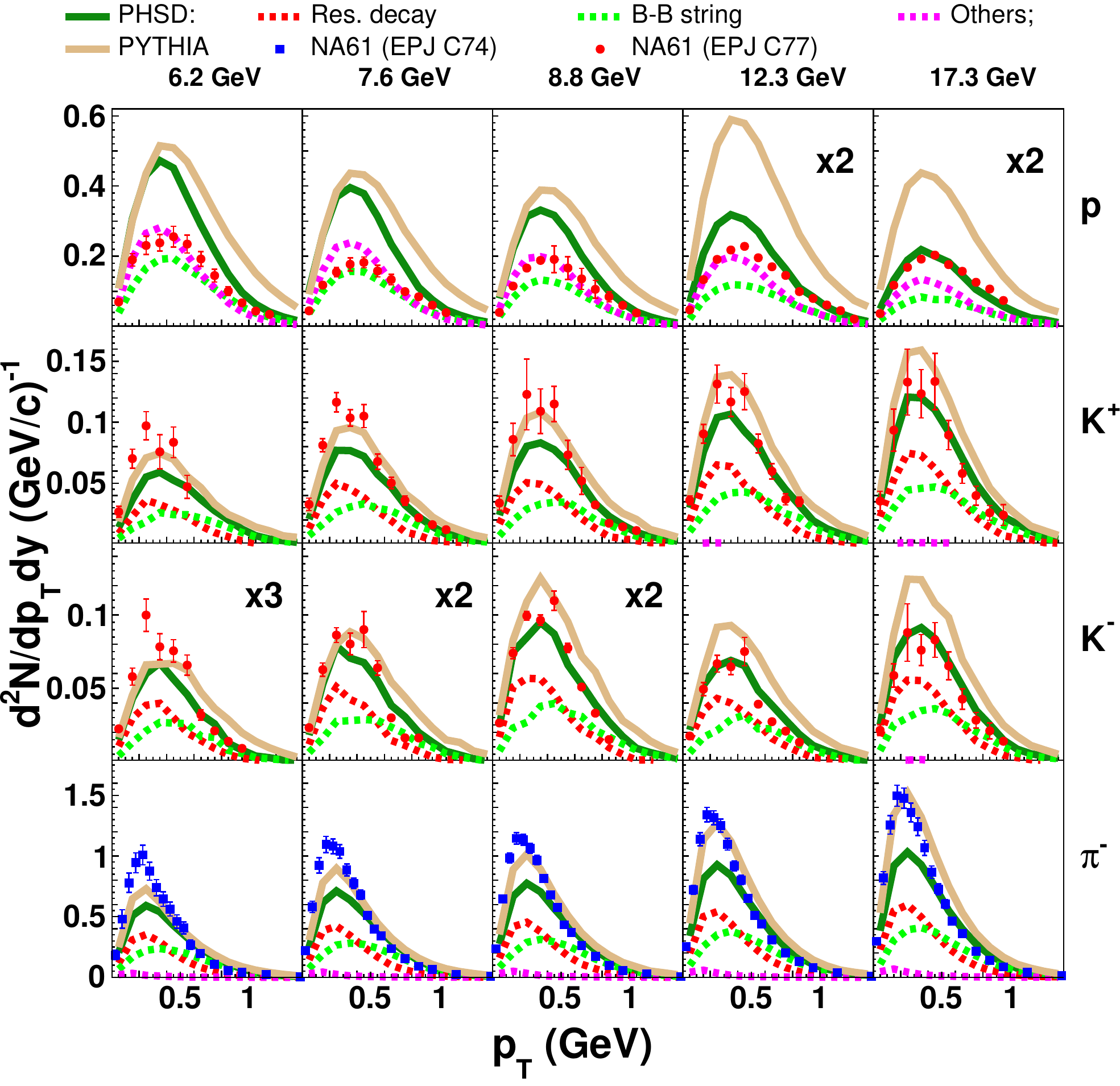}}
 \caption{Transverse momentum spectra of protons,  $K^+$, $K^-$ and 
 $\pi^-$ from $p+p$ collisions  
in the central midrapidity interval $0<y<0.2$ 
at 6.2 GeV, 7.6 GeV, 8.8 GeV, 12.3 GeV, 17.3 GeV.
The PYTHIA 8.2 results are plotted by brown solid lines, the PHSD results 
are presented by green solid lines. The channel decomposition of the PHSD results
are also shown: the contribution from the hadrons coming directly 
from the decays of $BB$ strings is plotted by light green dash-dotted lines while
those coming from baryonic or mesonic resonance decays 
are drawn by red dash lines, the magenta lines show the sum distribution form 
"other" sources during the final state interaction.  
The experimental data  are taken  from the NA61/SHINE Collaboration \cite{na61_hadrons}.
The scaling factors  for the data and theoretical results 
are introduced for better visualization: 
 $K^- \times 3$ for 6.2, 7.6, 8.8 GeV;   \ 
$\bar p \times 3$  for 12.3, 17.3 GeV. 
}
        \label{pt_all}     
    \end{figure*}

    \begin{figure}[!htbp]
        \resizebox{0.49\textwidth}{!}{      
        \includegraphics{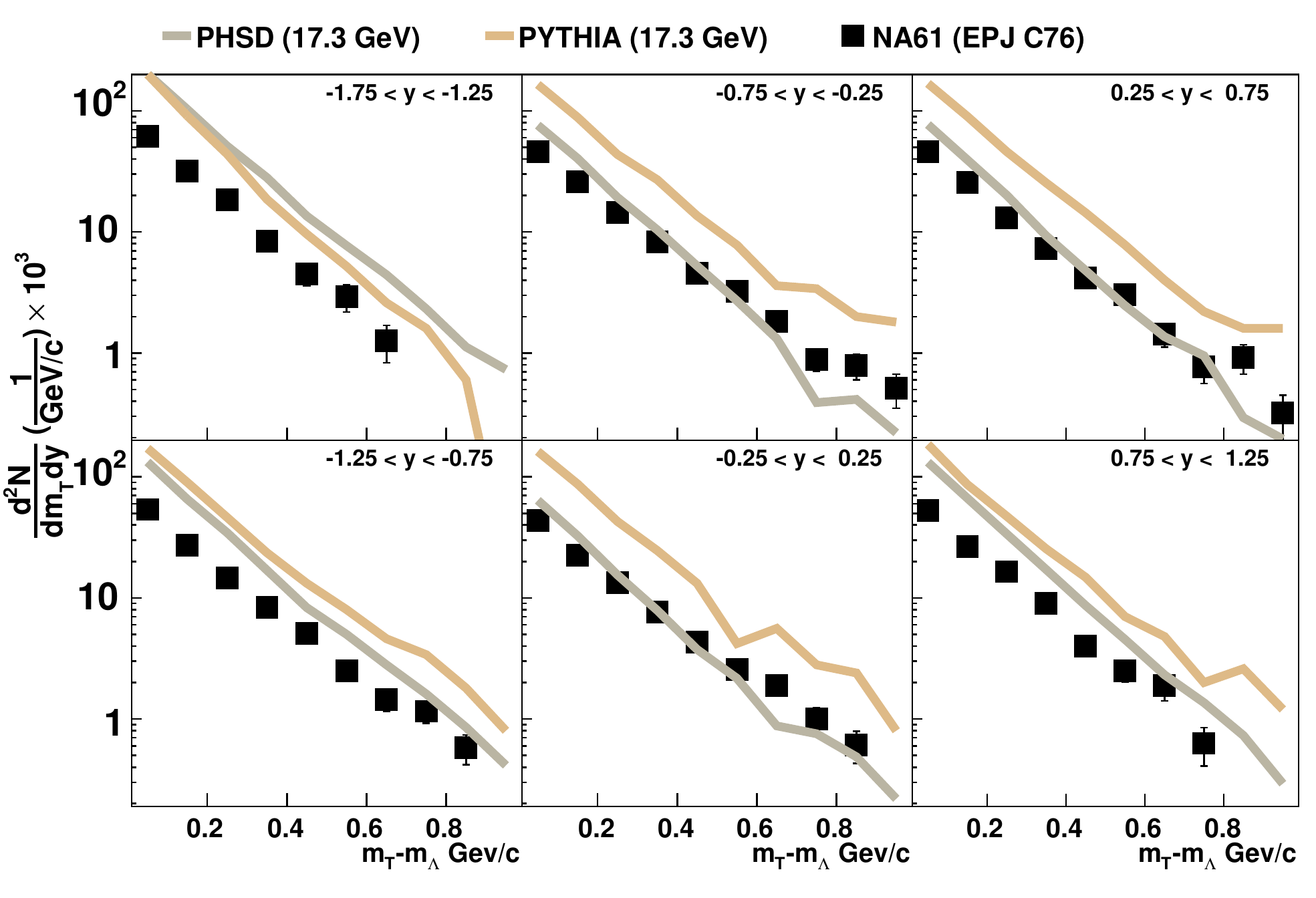}
      }
\caption{Transverse mass $m_T$ spectra of $\Lambda+\Sigma^{0}$ for different rapidity intervals ($-1.75\le y \le -1.25$, \ $-0.75\le y\le -0.25$, \ $0.25\le y\le 0.75$, \ 
$-1.25\le y\le -0.75$, \ $-0.25\le y\le 0.25$, \ $0.75\le y\le 1.25$)
from $p+p$ collisions at 17.3 GeV. 
The PYTHIA 8.2 results are plotted by brown solid lines, the PHSD results 
are presented by gray solid lines.
The experimental data  are taken  from the NA61/SHINE Collaboration \cite{na61_lambda158}. }
        \label{pt_lambdas}
    \end{figure}
    
Figure~\ref{pt_all} shows the transverse momentum 
distributions of protons, $\pi^-, K^\pm$ from inelastic $p+p$ collisions at 
$\sqrt{s_{NN}}=$ 6.2, 7.6, 8.8, 12.3, 17.3~GeV. The experimental data from 
the NA61/SHINE Collaboration~\cite{na61_hminus,na61_hadrons} are drawn by 
symbols, the spectra are measured near midrapidity ($0<y<0.2$). 
The PHSD results are plotted by green solid lines,  while the PYTHIA 8.2 by 
brown solid lines. We also show the contributions from different channels
for the PHSD spectra: the contribution from the hadrons coming directly 
from the decays of $BB$ strings is plotted by light green dash-dotted lines while
those coming from the baryonic or mesonic resonance decays 
are drawn by red dash lines, the magenta lines show the contribution form 
"other" sources during the final state interaction. As one can see the latter
is rather small for all hadron species. The hadrons stemming from string decay
show much harder spectra than from resonance decays which fill the low part 
of the $p_T$ distributions.

In Fig. \ref{pt_lambdas} we show the comparison of the PHSD and PYTHIA results 
for the transverse mass $m_T$ spectra for the strange baryons $\Lambda+\Sigma^0$ for different rapidity intervals at 
$\sqrt{s_{NN}}=$17.3 GeV in comparison to the data from the NA61/SHINE Collaboration
\cite{na61_lambda158}. In spite that the absolute values of the $m_T$ spectra are 
overestimated by PYTHIA 8.2 for all rapidity bins (for the reasons discussed in Section 4.4),
the slope of the theoretical spectra is approximately in line with the experimental data;
the PYTHIA slopes are slightly harder than the PHSD slopes.

\subsection{Excitation function of the inverse slope parameter of the 
$m_T-$ spectra of $K^\pm$ mesons}

    \begin{figure}[!htbp]
      \centering
      \resizebox{3.5in}{!}{
          \includegraphics{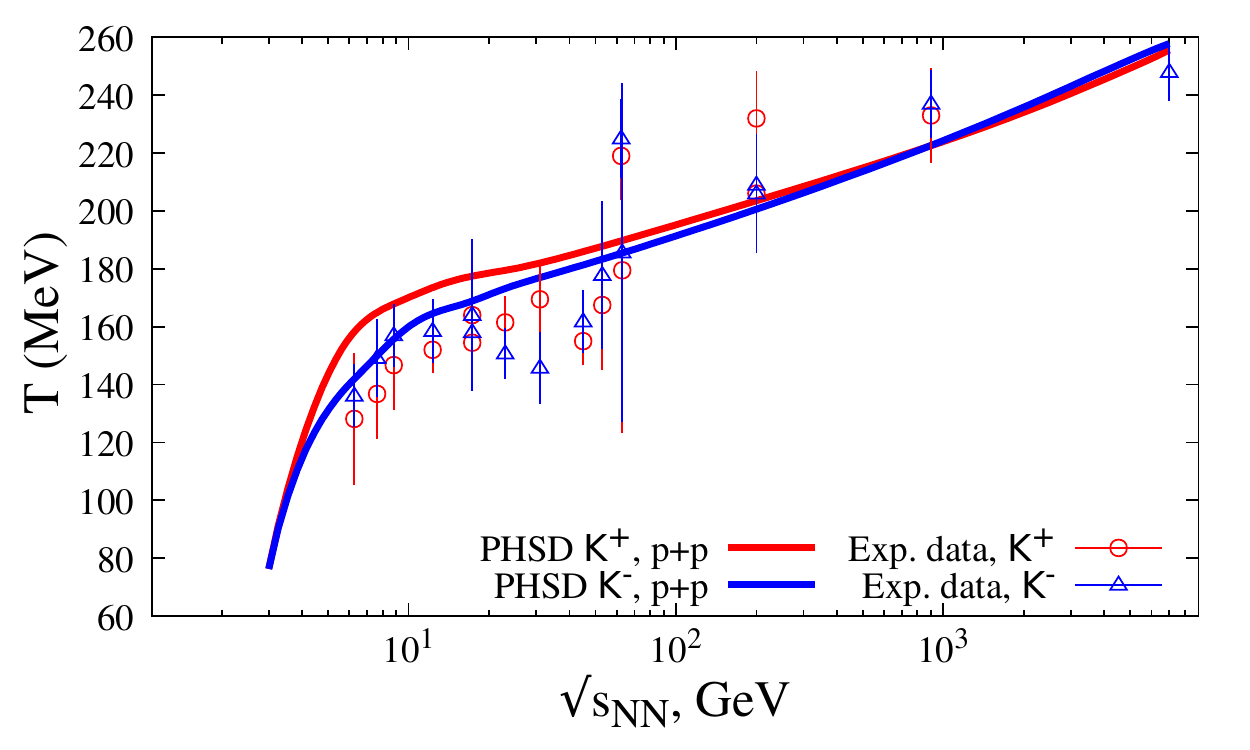}
      }
\caption{The inverse slope parameter $T$ of the $m_T$- spectra of $K^{\pm}$ mesons 
at mid-rapidity within the PHSD model.
 The compilation of experimental data are taken from \cite{Aduszkiewicz:2019zsv}.}
      \label{phys_slope}
    \end{figure}
    
Strangeness production in $A+A$ and $p+p$ collisions are always in the focus of the theoretical and experimental interest: 
the measured inverse slope $T$ of the $m_T$ spectra of $K^\pm$ mesons defined as 
\begin{eqnarray}
\label{slope}
\frac{1}{m_T} \frac{dN}{dm_T} \sim \exp(-\frac{m_T}{T})
\end{eqnarray}
shows a "step" behaviour in central $A+A$ collisions from 20 to 160 $A\cdot$GeV energies. 
This substantial flattening of the spectra in central Au+Au collisions relative to $p+p$
interactions has been  attributed to the onset of a deconfinement transition 
from hadronic to partonic matter \cite{Alt:2003rn,Gorenstein:2003cu}. 
As has been shown in Ref. \cite{Bratkovskaya:2003ie} such collective behaviour could 
not be reproduced by hadron based models (as HSD or UrQMD)
and might indicate the creation of pressure by partonic interactions in HICs \cite{Cassing:2008sv}.

In the last decade the experimental knowledge on the $m_T$ spectra of $K^\pm$ in $p+p$
collisions has been improved. Thus, we update and extend our previous study of the inverse
slope parameter $T$ of the midrapidity $m_T$ spectra of $K^\pm$ mesons 
(cf. \cite{Bratkovskaya:2003ie}) and present in Fig. \ref{phys_slope} the PHSD result
for the excitation function of $T$ versus $\sqrt{s_{NN}}$.
The compilation of the worldwide experimental data are taken 
from \cite{Aduszkiewicz:2019zsv}.
One can see that PHSD reproduces the $K^\pm$ slope rather well in a very wide energy range from a few GeV to a few TeV.


\subsection{Comparison of $y$- and $p_T$- distributions at $\sqrt{s_{NN}} = 200$ GeV}

    \begin{figure*}[!htbp]
      \centering
      \resizebox{0.7\textwidth}{!}{
        \begin{tabular}{cc}
            \includegraphics[page=1]{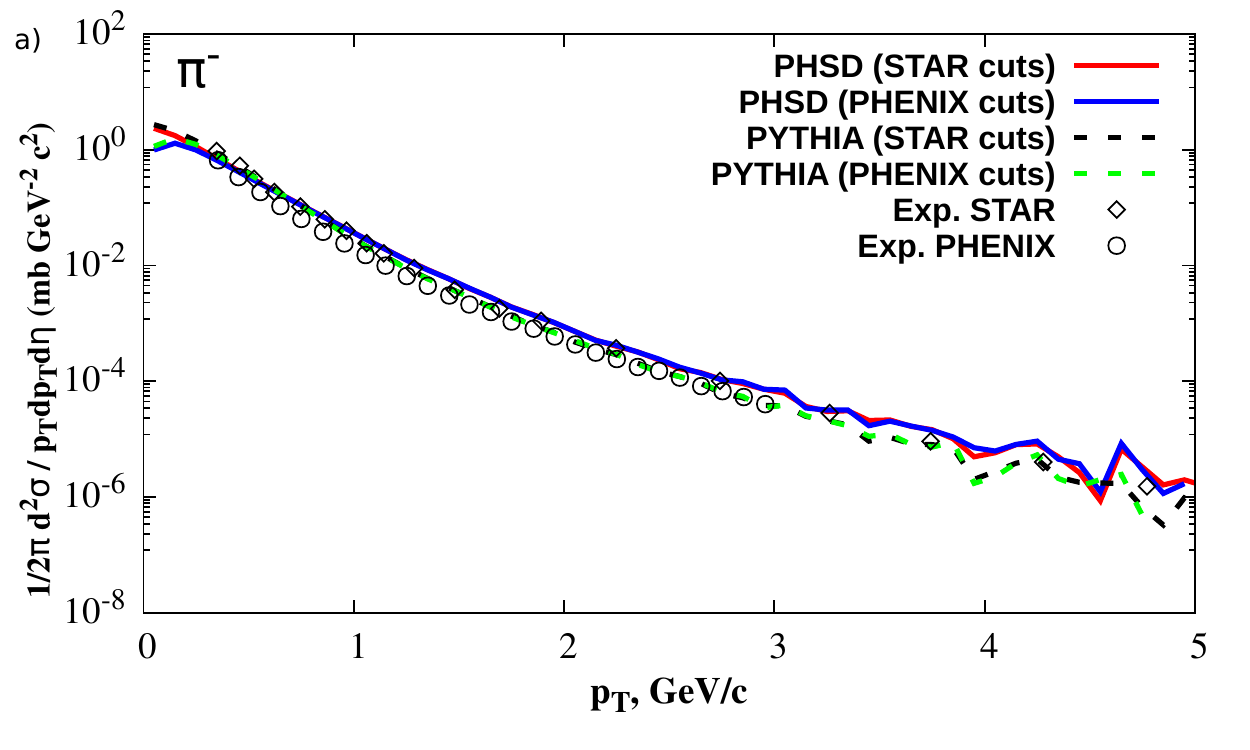} &
            \includegraphics[page=2]{phqmd_rhic_FSI.pdf} \\
            \includegraphics[page=3]{phqmd_rhic_FSI.pdf} &
            \includegraphics[page=4]{phqmd_rhic_FSI.pdf} \\
            \includegraphics[page=5]{phqmd_rhic_FSI.pdf} &
            \includegraphics[page=6]{phqmd_rhic_FSI.pdf} \\
            \includegraphics[page=7]{phqmd_rhic_FSI.pdf} &
            \includegraphics[page=8]{phqmd_rhic_FSI.pdf} \\
        \end{tabular}
      }
\caption{The invariant yields of  $\pi^{+}$, $K^{+}$ $p$, $\Lambda+\Sigma^{0}$
and $\bar\Lambda+\bar\Sigma^{0}$  as a function of $p_{T}$.
The cuts $|y| < 0.5$ and $|\eta| < 0.35$ were applied to the models for a comparison with 
STAR \cite{Abelev:2006cs} (open rhombus) and PHENIX \cite{Adare:2011vy} (open circles) 
data accordingly.
The PHSD results with STAR $y$-cut are shown as the red solid lines, with the PHENIX 
$\eta$-cut as blue solid lines. 
The PYTHIA 8.2 results with STAR $y$-cut are shown by the black dashed 
lines, with the PHENIX $\eta$-cut by green dashed lines.
}
      \label{rhic}
    \end{figure*}

    \begin{figure*}[!htbp]
      \centering
      \resizebox{0.8\textwidth}{!}{
        \begin{tabular}{cc}
          \includegraphics{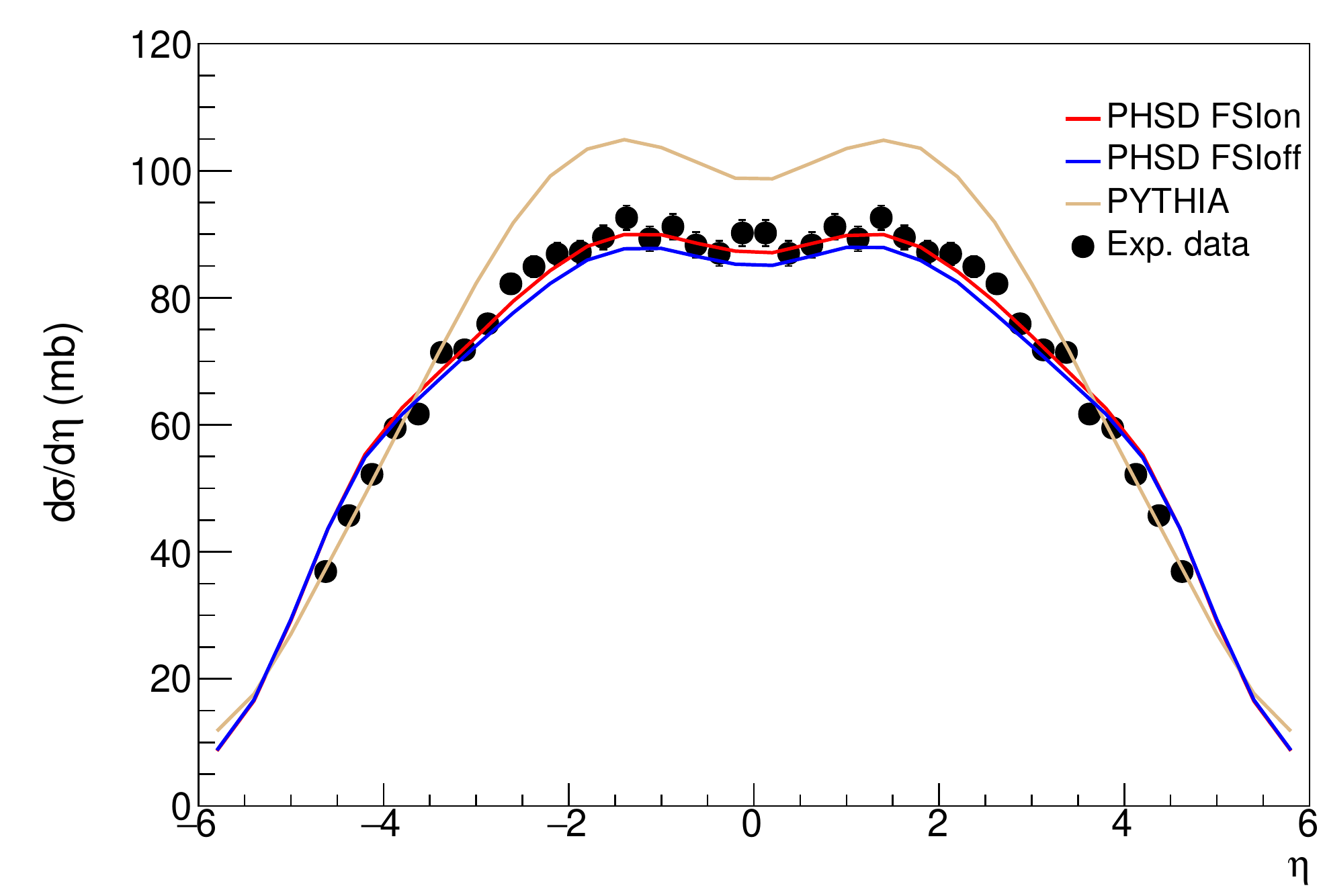}        
          \includegraphics{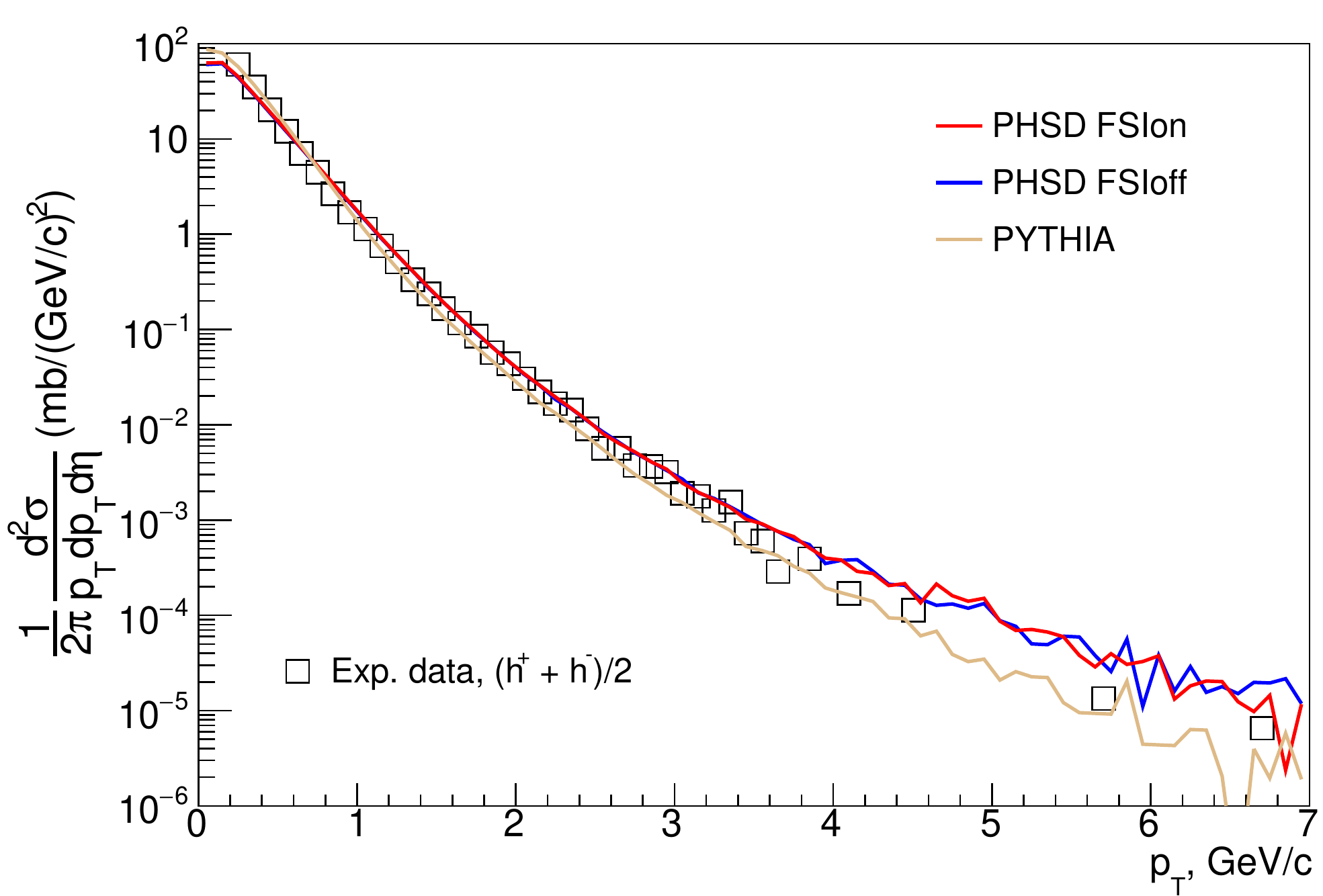}
        \end{tabular}
      }
\caption{
Left: The differential cross section $d\sigma/d\eta$ of negatively-charged hadrons
versus pseudorapidity $\eta$.
Right: The invariant cross-section of charged hadrons for $|\eta| < 2.5$
versus $p_T$. The PYTHIA 8.2 results are displayed by the brown lines.
The PHSD results are shown for two cases: including the FSI (default) 
by the red lines ('PHSD-FSIon') and without FSI by the blue lines 
('PHSD-FSIoff').
The experimental data from the UA5 and UA1 Collaborations are taken from 
Refs. \cite{Alner:1986xu,Albajar:1989an}.}
      \label{u5_ana}
    \end{figure*}

Now we step to ultra-relativistic energies and compare the PHSD and PYTHIA 8.2
results with data from the STAR  \cite{Abelev:2006cs} and PHENIX \cite{Adare:2011vy}
Collaborations in Fig. \ref{rhic}.
One can see that the experimental data on meson spectra are rather well reproduced 
by both models while the spectra of $\Lambda+\Sigma^{0}$ are slightly
underestimated and $\bar\Lambda+\bar\Sigma^{0}$ spectra are overestimated.
The agreement between PHSD and PYTHIA is quite good except of the low $p_T$ region 
for the baryons where the PYTHIA spectra are higher than the PHSD ones.

The latter is also observed in the differential cross section $d\sigma/d\eta$
of negatively-charged hadrons versus pseudorapidity $\eta$ 
presented in the left part of Fig. \ref{u5_ana}. Here the PYTHIA result overestimates 
the experimental data from the UA5 Collaboration \cite{Alner:1986xu} at mid-$\eta$
while  PHSD agrees very well with data. 
In Fig. \ref{u5_ana} the PHSD results are shown for two cases: including the FSI
(default for this study) by the red lines ('PHSD-FSIon') and without FSI   
by the blue lines ('PHSD-FSIoff'). 
The right part of  Fig. \ref{u5_ana} presents the invariant cross-section 
of charged particles for $|\eta| < 2.5$ versus $p_T$. One sees that
the $p_T$ spectra from the PHSD are harder at large $p_T$ and slightly softer 
at very low $p_T$. The latter can not be attributed to the FSI during the 
expansion rather than to the differences in string fragmentation since 
the PHSD results  with and without FSI are very close to each other in the whole
$p_T$ range. The FSI leads to a very small enhancement of the hadron multiplicity
at mid-rapidity which has been also shown in Fig. \ref{ratio_fsi}.

\subsection{Comparison of $p_T$ spectra at LHC energies, traces of the final state interactions}

    \begin{figure}[!htbp]
      \centering    
      \resizebox{0.45\textwidth}{!}{      
          \includegraphics{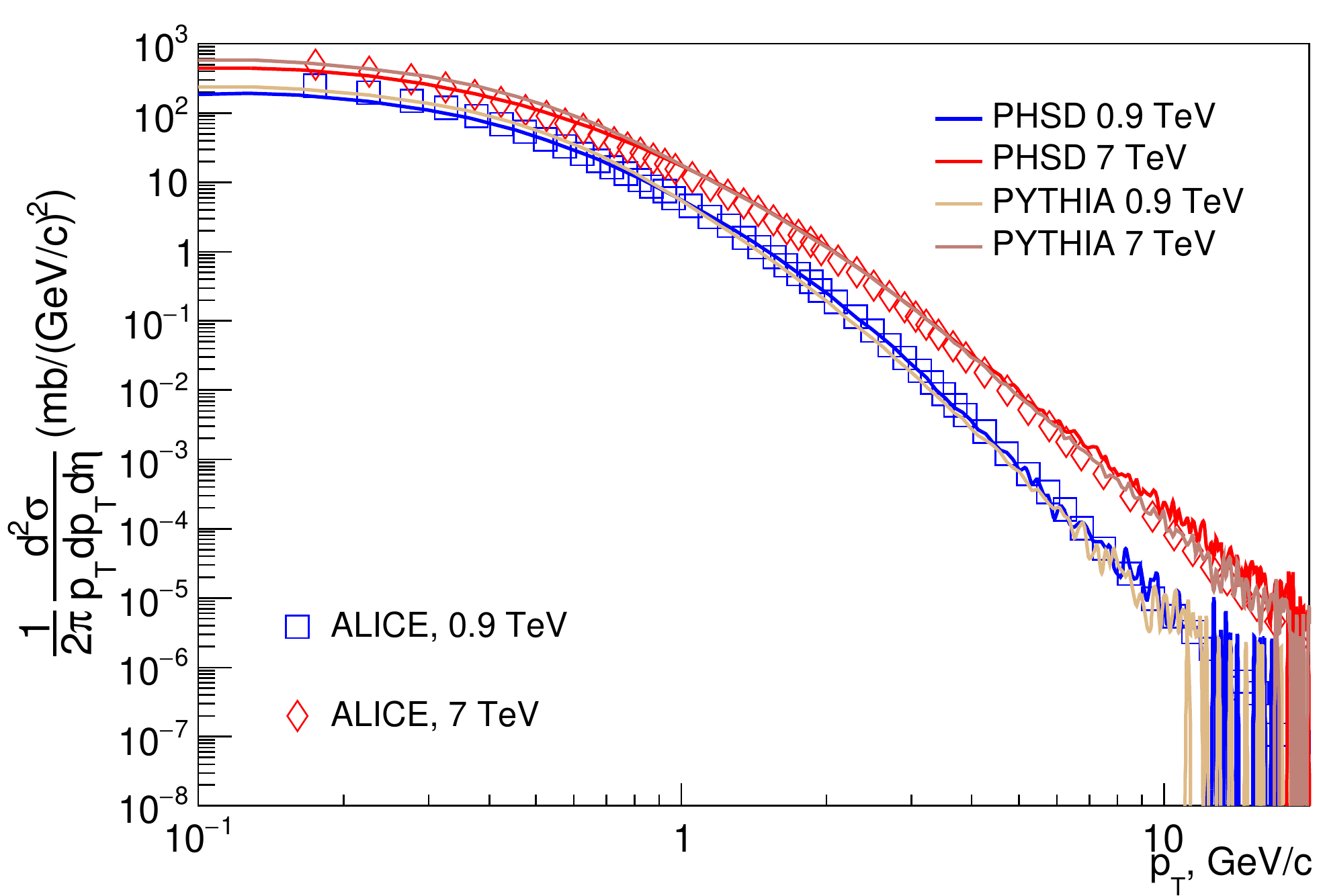}
      }
\caption{The invariant cross section versus $p_T$ for charged particles for $|\eta| < 0.8$
for $p+p$ collisions at invariant energies $\sqrt{s_{NN}}=$0.9 TeV and 7 TeV.
The PHSD results are indicated by the blue line for 0.9 TeV and by the red line for 7 TeV.
The PYTHIA 8.2 results are shown by the brown line for 0.9 TeV and by the grey line 
for 7 TeV.  The experimental data from the ALICE Collaboration  \cite{Abelev:2013ala} 
at 0.9 TeV are shown as open rhombus and at 7 TeV as open squares.      
      }
      \label{lhc_alice}
    \end{figure}

We increase in energy up to the LHC now and come to a comparison of the 
PHSD and PYTHIA 8.2 results to the ALICE data.
In Fig. \ref{lhc_alice} the invariant cross section versus $p_T$ for charged particles for
$|\eta| < 0.8$ for $p+p$ collisions at invariant energies $\sqrt{s_{NN}}=$0.9 TeV and 7 TeV
are shown. The PHSD results are indicated by the blue line for 0.9 TeV and by the 
red line for 7 TeV. 
The PYTHIA 8.2 results are shown by the brown line for 0.9 TeV and by grey line for 7 TeV.      
The experimental data from the ALICE Collaboration  \cite{Abelev:2013ala} 
at 0.9 TeV are shown as open rhombus and at 7 TeV as open squares.       
One can see that the PHSD and PYTHIA $p_T$ distributions have a similar slope, 
however, the PYTHIA spectra are slightly higher. Both models are in a good agreement with ALICE data which cover 10 orders of magnitude in range.


    \begin{figure*}[!htbp]
      \centering
      \resizebox{\textwidth}{!}{
        \begin{tabular}{ccc}
            \includegraphics[scale=1.5]{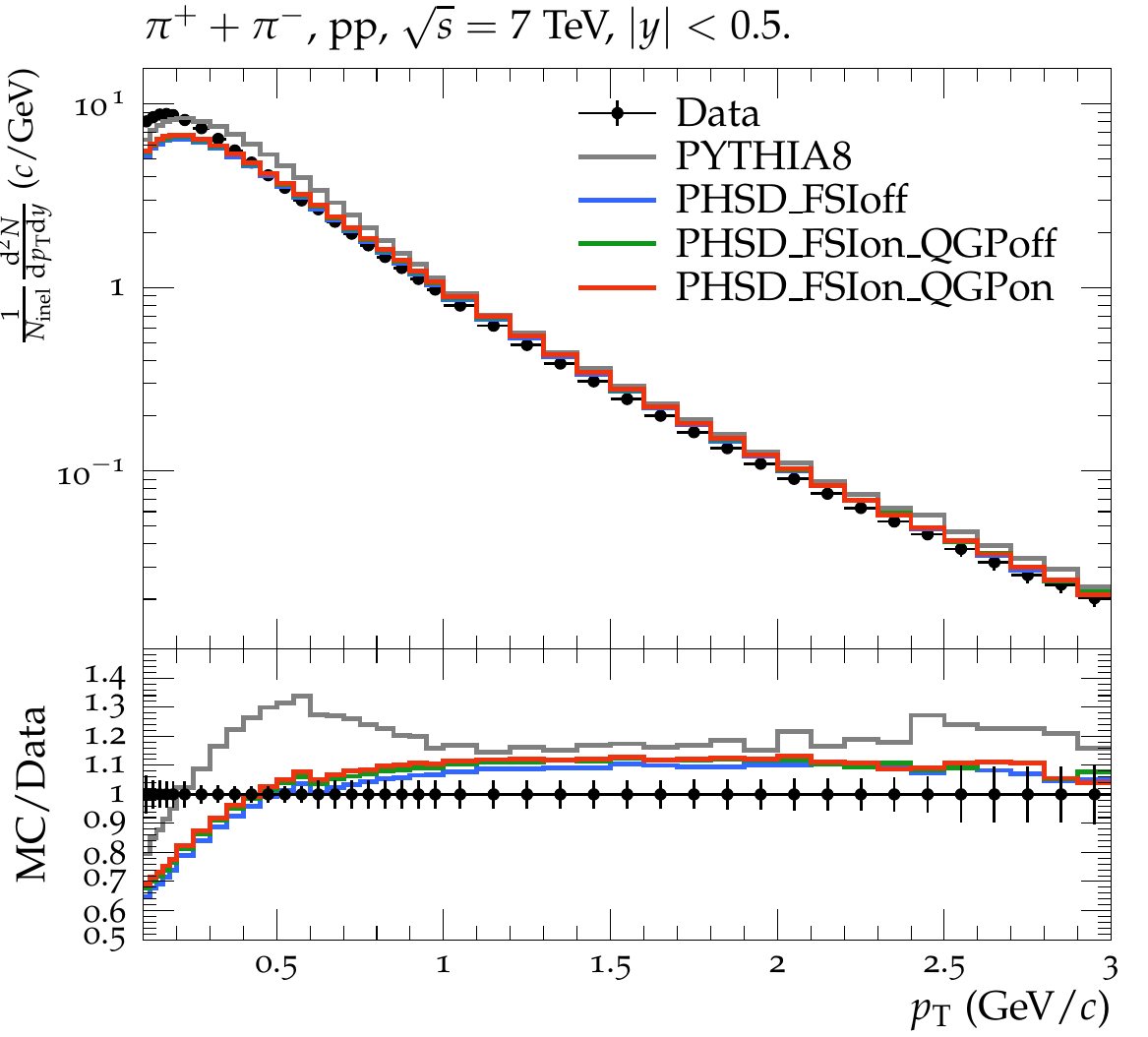}  &
            \includegraphics[scale=1.5]{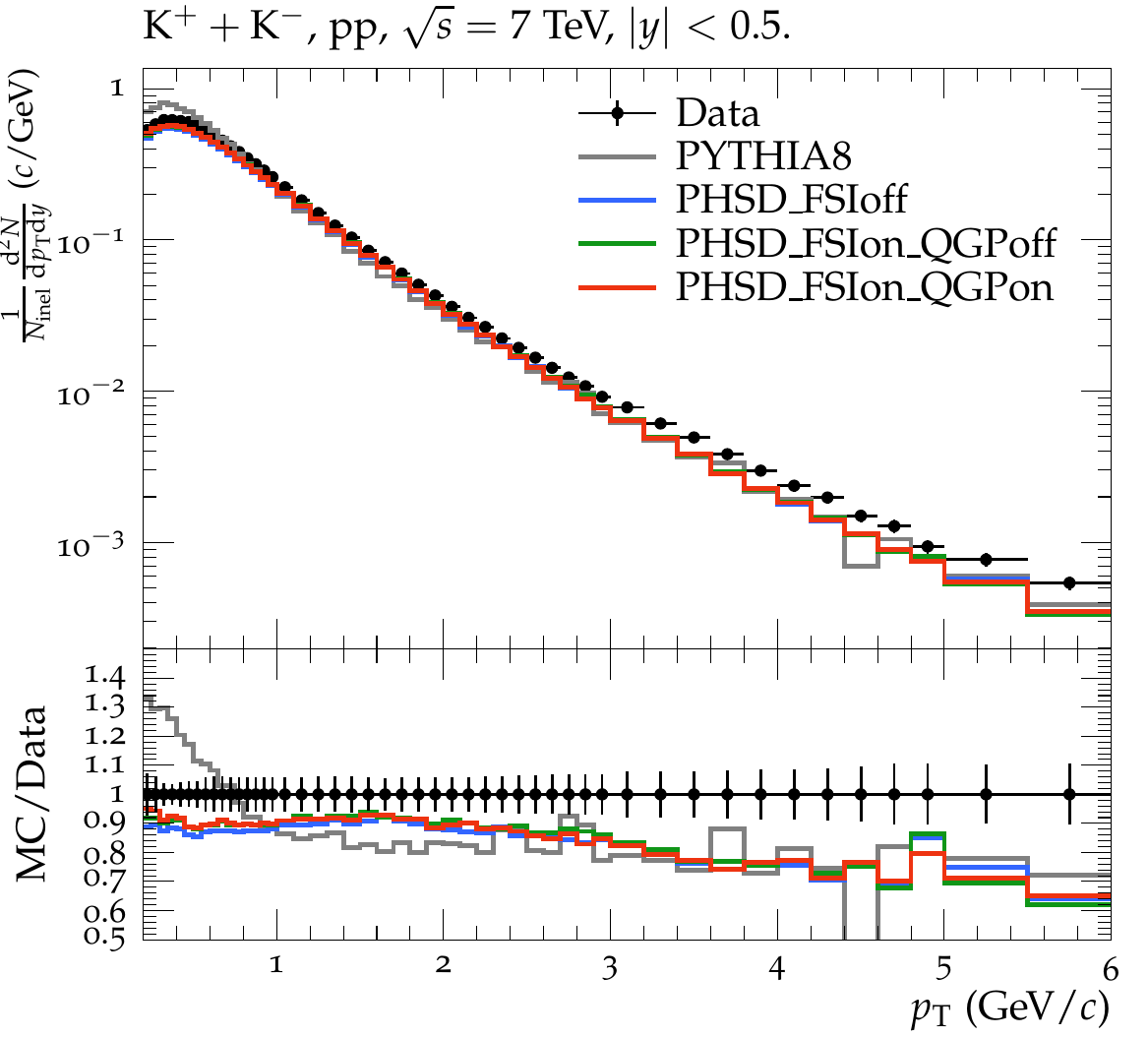}  &
            \includegraphics[scale=1.5]{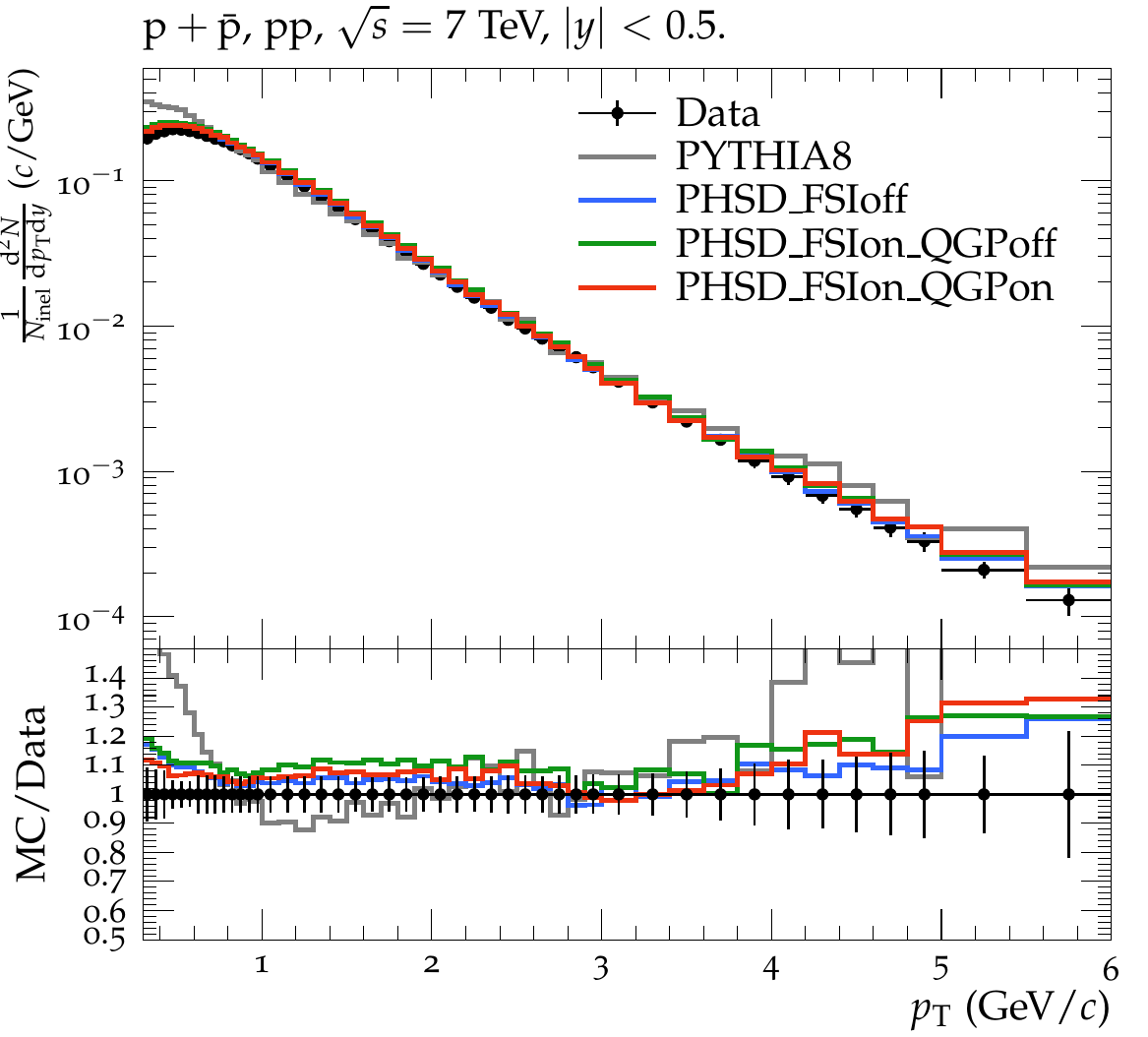} 
        \end{tabular} }\\
       Ratio FSIon/FSIoff, all $N_{ch}$
      \resizebox{\textwidth}{!}{        
        \begin{tabular}{ccc}
            \includegraphics[scale=1.5]{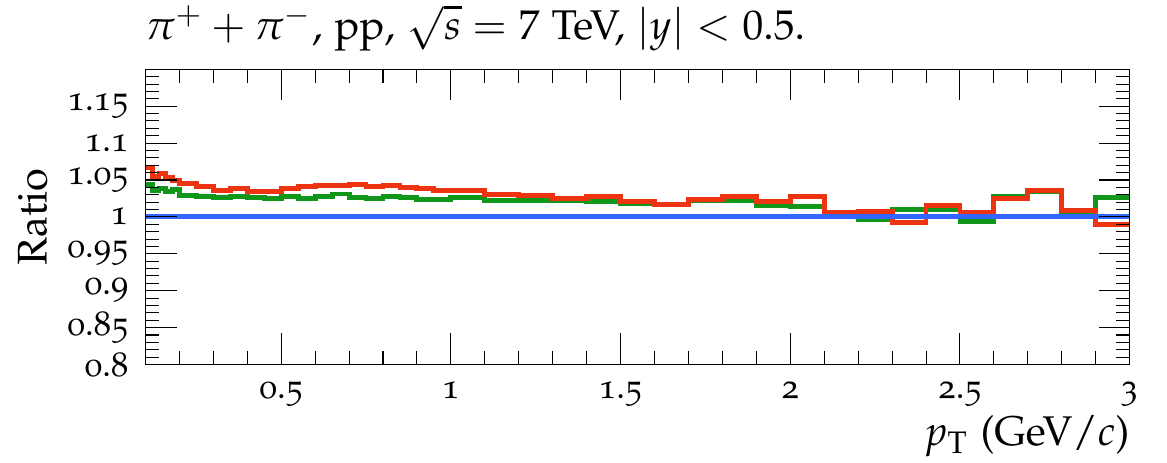}  &
            \includegraphics[scale=1.5]{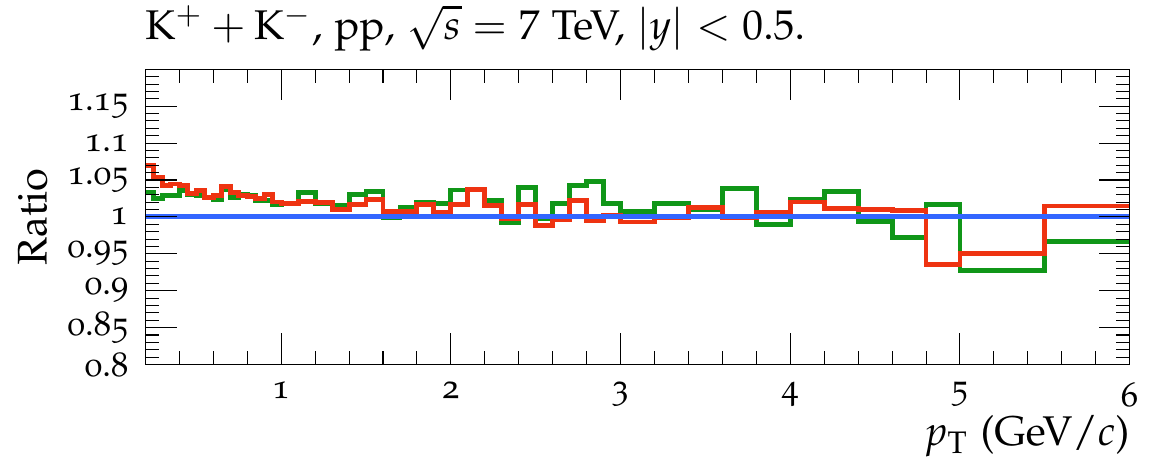}  &
            \includegraphics[scale=1.5]{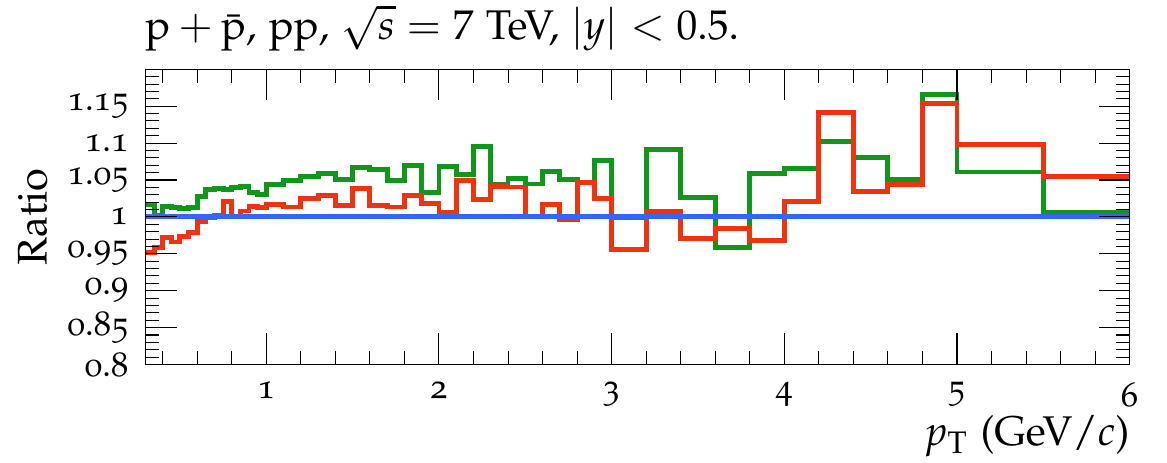} 
        \end{tabular}  }\\
         Ratio FSIon/FSIoff, $N_{ch}>80$
      \resizebox{\textwidth}{!}{        
        \begin{tabular}{ccc}
            \includegraphics[scale=1.5]{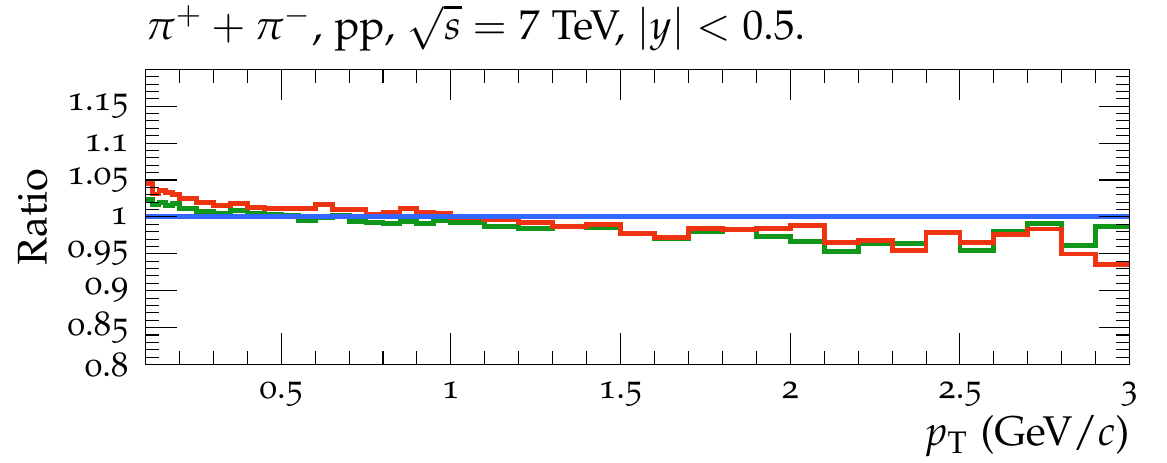}  &
            \includegraphics[scale=1.5]{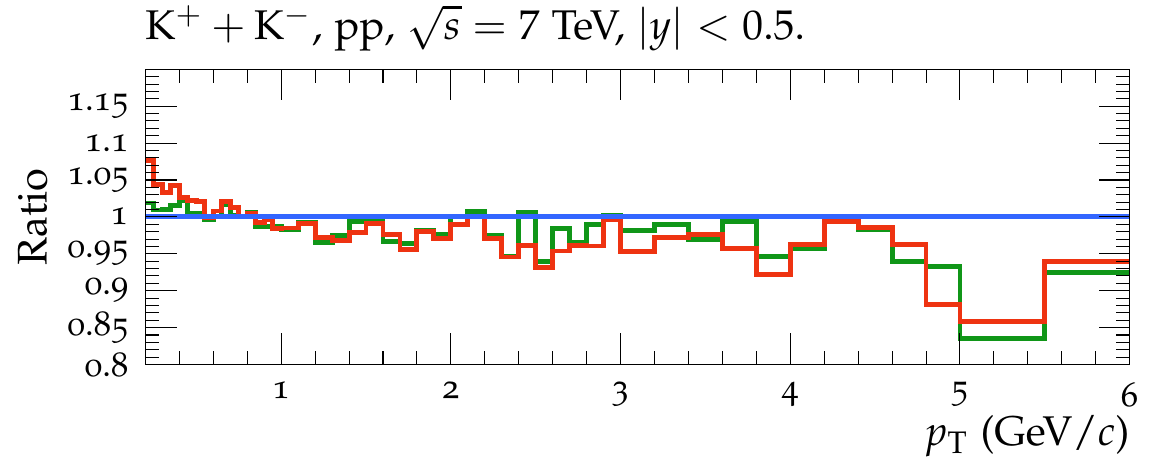}  &
            \includegraphics[scale=1.5]{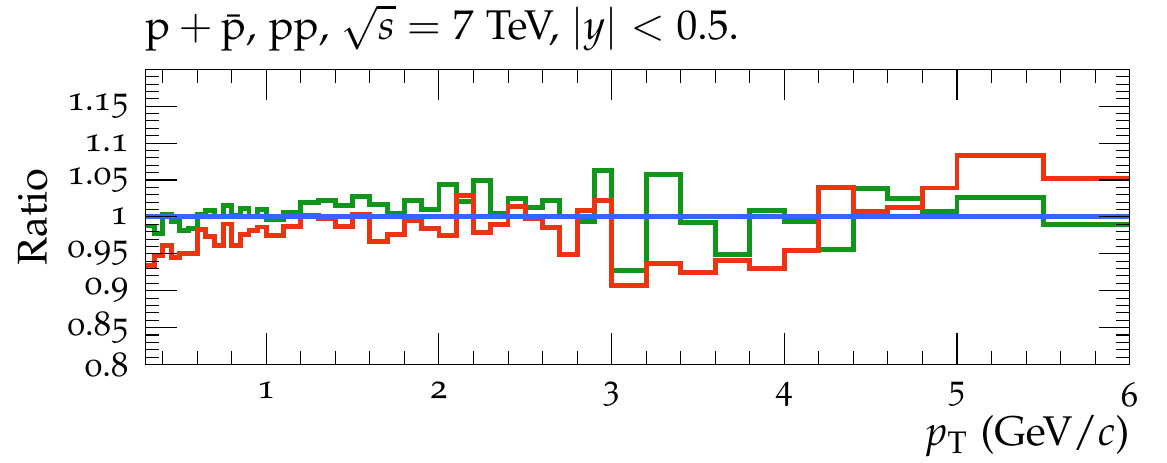} 
        \end{tabular}              
      }
\caption{Upper row:
The transverse momentum spectra of $\pi^{+}+\pi^-$ (left), $K^{+}+K^-$ (middle),
and $p+\bar p$ (right) in midrapidity ($|y| < 0.5$) $p+p$ collisions at 
$\sqrt{s_{NN}}=7$ TeV.
The grey lines correspond to the PYTHIA 8.2 results, the blue lines to the PHSD results 
without FSI ('PHSD-FSIoff'), the green lines to the PHSD results 
with hadronic FSI, but without QGP creation ('PHSD-FSIon-QGPoff'), 
the red lines to the PHSD results with hadronic FSI and with QGP creation 
('PHSD-FSIon-QGPon'). 
The solid dots indicate the experimental data from the ALICE Collaboration 
\cite{Adam:2015qaa}. The deviation of the model results from the data 
are shown directly under each plot. 
Middle row: the ratio of the PHSD transverse momentum spectra (from the upper plots)
calculated with hadronic FSI but without QGP creation to the spectra without FSI 
(FSIon-QGPoff/FSIoff) shown by  green lines,  
with FSI with QGP to the spectra without FSI (FSIon-QGPon/FSIoff) by red lines.
Lower row: the same as middle, but for number of charge particles $N_{ch}>80$.
The analysis is performed using Rivet \cite{Buckley:2010ar}.
}
      \label{pt7TeV}
    \end{figure*}

We continue with a model comparison of transverse momentum spectra 
of identified hadrons in $p+p$ collisions to the spectra measured 
by the ALICE Collaboration  \cite{Adam:2015qaa} at 7 TeV. 
At such ultra-relativistic energy a large amount of hadrons are produced
during the string breaking which leads to large energy-density fluctuations 
and to the possible creation of small droplets of QGP especially in 
the events with very high multiplicities.
The experimental observation of a visible $v_2$ (which is even comparable 
with the $v_2$ of heavy-ions) in high multiplicity $p+p$ collisions \cite{Aaboud:2016yar} 
indicates the development of collective effects (i.e. hydrodynamic behaviour) 
in such small system \cite{EPOSLHC,Shuryak:2013ke,Bzdak:2013zma} which 
might be also in line with the idea of QGP formation in high multiplicity $p+p$. 

In order to study the possible traces of the QGP formation on hadronic 
'bulk' observables - as $p_T$ spectra - we perform PHSD calculations 
when additionally to the hadronic final state interactions (default), we
consider the formation of the QGP after the initial $pp$ string breaking 
in a similar way as in HICs (cf. Section 2). 
Indeed, the QGP formation in $p+p$ might happen only in a few cells
where, due to fluctuations, the local energy density becomes 
larger than the critical $\varepsilon_C \simeq 0.5$ GeV/fm$^3$ such that
a dissociation of hadrons to partons occurs in this cell. However, the size and
the life time of such QGP droplets are very small contrary to HICs,
they carry only a very small fraction of the total energy in the collisions, 
thus, one could not expect a larger effect of the QGP creation on bulk observables.

In Fig. \ref{pt7TeV} we show
the $p_T$ spectra of $\pi^{+}+\pi^-$ (left panel), $K^{+}+K^-$ (middle panel),
and $p+\bar p$ (right panel) in midrapidity ($|y| < 0.5$) $p+p$ collisions 
at $\sqrt{s_{NN}}=7$ TeV.
The grey lines correspond to the PYTHIA 8.2 results, the blue lines to the PHSD results 
without FSI ('PHSD-FSIoff'), the green lines  to the PHSD results 
with hadronic FSI, but without QGP creation ('PHSD-FSIon-QGPoff'), 
the red lines to the PHSD results with hadronic FSI and with QGP creation 
('PHSD-FSIon-QGPon'). The analysis is performed using Rivet \cite{Buckley:2010ar}
which allows to show the deviation of the models from the experimental 
data below each plot.
One can see that PYTHIA 8.2 creates more very low momentum hadrons
than the PHSD (which we attribute to the Innsbruck tune of string routines 
used in the PHSD). With increasing $p_T$ both models show a similar trend:
pion $\pi^{+}+\pi^-$ spectra are slightly harder in the models 
while $K^+ + K^-$ spectra are softer;
the $p+\bar p$ spectra agree very well with data up to 5 GeV/c. 

In order to quantify the role of the final state interactions 
we show additionally the ratios of the PHSD transverse momentum spectra 
(from the upper plots) calculated with hadronic FSI but without QGP creation 
to the corresponding spectra without FSI (FSIon-QGPoff/FSIoff) by the 
green lines in the middle row;  
with FSI with QGP to the spectra without FSI (FSIon-QGPon/FSIoff) by red lines.
The lower row indicates the same ratios as the middle row, but for the number of 
charged particles $N_{ch}>80$, i.e. by selecting the events with large multiplicities.
One can see from Fig. \ref{pt7TeV} that the FSI effect is relatively small, 
on the level of 5\% in average.
This PHSD result is consistent with a recent finding by Sj\"ostrand and Utheim \cite{Sjostrand:2020gyg} who incorporated the framework for accounting of 
the FSI in  PYTHIA in terms of hadronic rescattering.

As follows from  Fig. \ref{pt7TeV}, the PHSD calculations with
the hadronic FSI as well as with hadronic FSI and QGP creation lead 
to a small softening of the low $p_T$ pion and kaon spectra and hardening 
of proton+antiproton spectra. 
This is attributed to elastic scattering (which has a forward peaked angular 
distribution for $B+B$ and $m+B$ collisions) as well as to the inelastic processes 
and formation of resonances (dominantly $\Delta$'s).  
The high $p_T$ region is less sensitive to the FSI.

\subsection{Comparison of multi-strangeness production at LHC energies}

    \begin{figure}[!htbp]
      \centering    
      \resizebox{0.35\textwidth}{!}{      
          \includegraphics{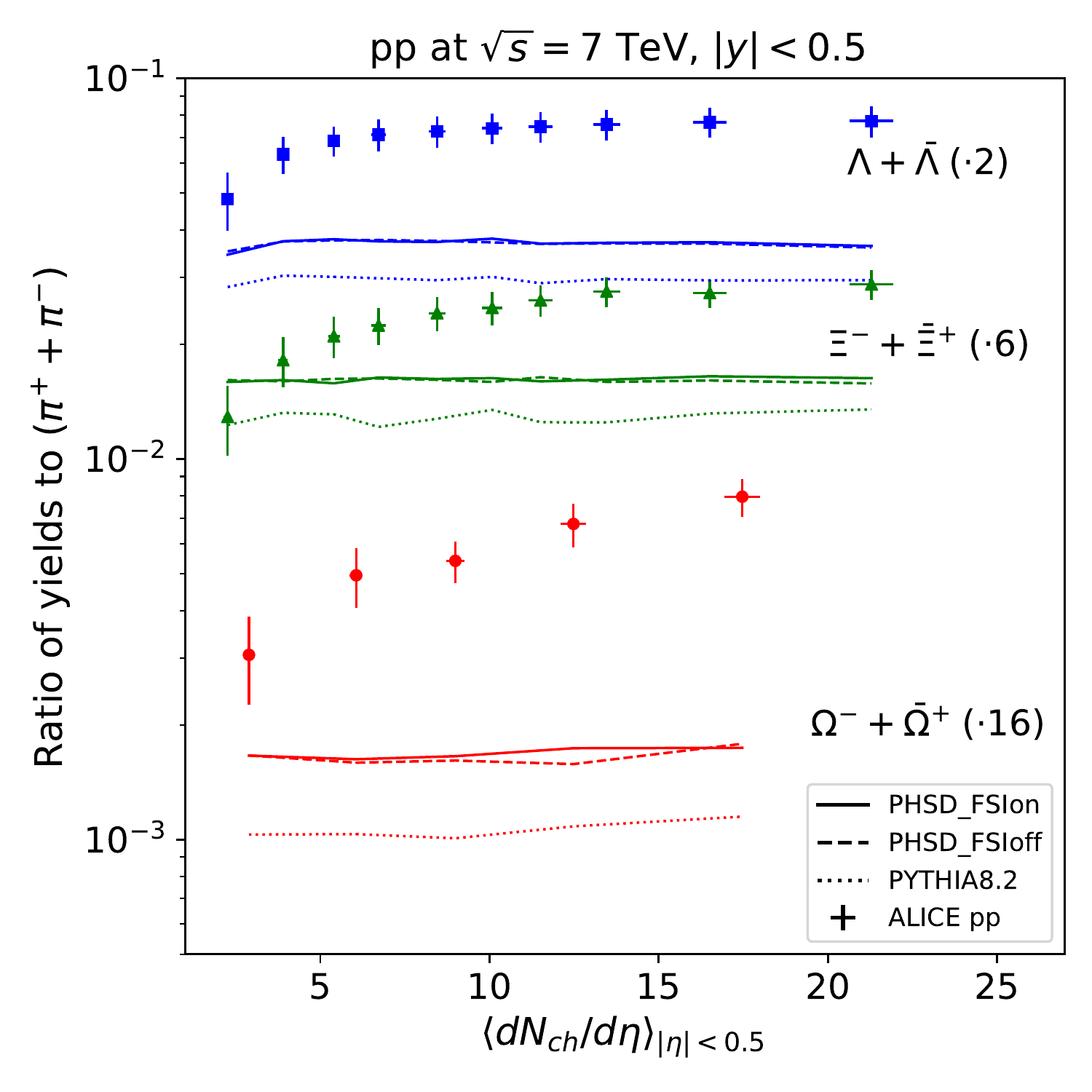}
      }
\caption{The ratio of the $p_T$-integrated yield of $\Lambda+\bar\Lambda$ (multiplied by a factor 2), $\Xi^- + \bar{\Xi}^{+}$ (multiplied by 6), $\Omega^- + \bar\Omega^+$
(multiplied by 16) to pions $(\pi^+ +\pi^-)$ as a function of 
$dN_{ch}/d\eta$ for $|y|<0.5$ at $\sqrt{s_{NN}}=7$ TeV. 
The dotted lines correspond to the PYTHIA 8.2 results, the dashed lines to the PHSD results 
without FSI ('PHSD-FSIoff'), the solid lines to the PHSD results with hadronic FSI 
('PHSD-FSIon').
The experimental data from ALICE Collaboration are taken from Ref. \cite{ALICE:2017jyt}.      
      }
      \label{lhc_ABpi}
    \end{figure}
    
Finally, we step to the multi-strangeness production in $pp$ reactions and compare in 
Fig. \ref{lhc_ABpi} the PHSD and PYTHIA 8.2 results for the ratio of the 
$p_T$-integrated yield of $\Lambda+\bar\Lambda$ (multiplied by a factor 2), 
$\Xi^- + \bar{\Xi}^{+}$ (multiplied by 6), 
$\Omega^- + \bar\Omega^+$ (multiplied by 16) to pions $(\pi^+ +\pi^-)$ as a function of 
$dN_{ch}/d\eta$ for $|y|<0.5$ at $\sqrt{s_{NN}}=7$ TeV with the ALICE data 
from Ref. \cite{ALICE:2017jyt} (we keep the multiplication factors as in 
Fig. 2 of Ref. \cite{ALICE:2017jyt} for easy comparison). 
We mention that the analysis has been performed using Rivet.
As seen from Fig. \ref{lhc_ABpi}, both models can not reproduce the 
enhancement of multi-strange baryon production compared to the non-strange hadrons 
in high multiplicity events as observed by the ALICE Collaboration. 
As follows from the PHSD results with and without the hadronic FSI, the final
rescattering on the hadronic level can not enhance the ratio since the "chemistry production"
is mainly attributed to the very initial stages of $pp$ collisions. 
The "QGP" scenario (we omit to show it explicitly in Fig.\ref{lhc_ABpi} since
it is similar to the other scenarios within statistical fluctuations) in PHSD also 
can not make it since (as explained above) the QGP
is formed by melting the "pre-hadrons" during the expansion phase
and the QGP droplets can be formed due to the fluctuations in energy density. 
Thus, the PHSD and PYTHIA 8.2 results are qualitatively close to each other, 
the differences between them can be attributed to the different settings 
for the strange diquark production. 

We note that the enhancement  of strange to non-strange hadron production 
in $pp$ reactions has been reproduced within the EPOS-LHC model \cite{EPOSLHC} by 
the collective hadro\-nization of hot 'core' which decays in a statistical way to hadrons  
as well as by the DIPSY Monte-Carlo event generator in Ref. \cite{Bierlich:2014xba} 
by introducing 'colour ropes'. Such mechanisms are not incorporated in the default 
PYTHIA 8.2 and in the PHSD.


\section{Summary}

We have studied the hadron production in $p+p$, $p+n$ and $n+n$ reactions within 
the PHSD which is a microscopic transport approach for the dynamical description
of $A+A$ and $p+A$ collisions and compared the PHSD results with PYTHIA 8.2. 
In the PHSD the time evolution of collisions is described by the solution of generalized
transport equations derived from the first-order gradient expansion of Kadanoff-Baym
equations applicable for strongly interacting systems. 
In the PHSD all interactions in the system - 
on a hadronic or partonic levels - are treated in a fully microscopic way.
The multiparticle production from the primary energetic $NN$ collisions 
as well as from secondary $BB$, $mB$ and $mm$ reactions are based on the Lund string model
realized in terms of the event generators FRITIOF and PYTHIA.

The Lund event generators FRITIOF and PYTHIA have been developed with the focus 
on elementary reactions at ultra-relativistic energies. 
However, the FRITIOF and PYTHIA generators are very important also for the 
description of heavy-ion physics since they historically have been incorporated 
in many transport approaches.
Such applications to HICs requires from elementary event generators a good description 
of $BB$, $mB$ and $mm$ reactions in very wide energy range - from few GeV to a few TeV. 
Moreover, the flavour "chemistry" of elementary reactions, happening during 
the time evolution of  HICs, covers all possible flavour combinations of the 
colliding hadrons.
Additionally, the $hh$ interactions in HICs are happening in a hot and dense 
environment and not in vacuum as in  "free" $p+p$ collisions.
This requires  a modification ("tune") of the original Lund string model
which we have presented here within the PHSD approach.

The {\bf "PHSD tune"} of the Lund string model (FRI\-TIOF 7.02 and PYTHIA 6.4 generators) 
contains of few basic directions which could be summarized as\\
{\bf I)} an improvement of the description for the elementary reactions in the vacuum: 
\begin{itemize}
\item an extension of the applicability range of the Lund generators to very low energies
\item an improvement on the flavour "chemistry" of produced hadrons
\item a  modification of the string fragmentation function, i.e. in energy-momentum 
distributions for a better description of low energy data on hadron production
\end{itemize}
{\bf II)} a modification of string fragmentation and the properties of produced hadrons
in the hot and dense medium created in HICs:
\begin{itemize}
\item an implementation of  chiral symmetry restoration via the Schwinger mechanism 
for string decay in the dense medium
\item accounting for the initial state Cronin effect for $<k_T>$ broadening in the medium
\item implementation of the in-medium properties of hadrons in the string fragmentation
by incorporation of the in-medium spectral functions for mesonic and baryonic resonances
with momentum, density and temperature dependent widths instead of
non-relativistic spectral functions with constant width.
\end{itemize}
{\bf III)} We also pointed out the conceptual difference in the treatment of free 
(i.e. in the vacuum) $N+N$ collisions between the PHSD and PYTHIA models.
In the default PYTHIA 8.2 the hadrons are produced by the string fragmentation
which provides the momenta of outgoing particles, however, the space-time picture
of $p+p$ collisions is not presented here.
In the PHSD the free $N+N$ collisions are treated in a similar fashion as in $A+A$,
i.e. following the space-time and momentum evolution of the system by solving 
the relativistic transport equations of motion. Moreover, the hadrons produced 
from primary string fragmentation can participate in the final state interactions 
by hadronic rescattering.
Furthermore, at ultra-relativistic collisions  small droplets of  QGP 
could be formed in events with a high multiplicity of produced hadrons due to
energy-density fluctuations.

In this study we have presented a detailed comparison of the PHSD results 
with those from the default version of PYTHIA 8.2 for 'bulk' observables 
such as the excitation functions of hadron 
multiplicities as well as differential rapidity $y$, transverse momentum $p_T$
and $x_F$ distributions in $p+p$, $p+n$ and $n+n$ reactions in the energy range 
$\sqrt{s_{NN}} = 2.7 - 7000$ GeV where we also compared the models with the existing 
experimental data. 

We found that 
i) in general the extrapolation of the Lund model 
(realized by the FRITIOF, PYTHIA generators) to low energies (much below the default threshold) works rather well for the description of total multiplicities of produced 
hadrons which validates its use as elementary event generators in transport approaches. 
However, some tuning is still required; the experimental data on the multiplicities 
of produced hadrons at low and intermediate energies are better described with
the "tuned strings" in PHSD. The same holds for the differential observables
as rapidity and $p_T$ spectra.
However, a further improvement of the string fragmentation is required in order 
to obtain a better description of experimental data at low and intermediate energies, especially for the production of multi-(anti-)strange hadrons.

ii) We showed a strong isospin dependence of particle production in 
$p+p$, $p+n$ and $n+n$ reactions, especially at low energies. 
However, the lack of experimental data doesn't allow to make  reliable constrains here.
In this respect experimental data on proton + light nuclei collisions 
might be helpful. 

iii) We have investigated the role of final state interactions due to the hadronic
rescattering on the bulk observables and found that at low energies it is negligible 
due to a very low density of produced hadrons; the FSI effect grows with 
increasing collision energies, however, even at the LHC energies it gives less 
then 5\% increase of the charged hadron multiplicities and only small changes 
in the transverse momentum spectra. 
This PHSD finding is in line with the recent results by the Lund group 
\cite{Sjostrand:2020gyg} where the hadronic FSI effect has been incorporated
in PYTHIA  within a framework of the space-time picture of $p+p$ collisions.
We also showed the influence on $p_T$ spectra of the possible small QGP droplet formation 
in $p+p$ collisions at LHC energies and found only a very small effect here.

Finally, we stress the importance of the development of reliable event generators 
for elementary reactions from low to ultra-relativistic energies in view 
of heavy-ion physics.

\section*{Acknowledgements}
The authors acknowledge inspiring discussions and useful remarks 
from  T.~Sj\"ostrand and C. Bierlich.
They are also grateful to C. Blume, W. Cassing,  V. Lenivenko, P. Moreau, L. Oliva, 
 K. Shtejer, O. Soloveva, T. Song and K. Werner for useful discussions and their 
interest to our work.  We are grateful to S. Pulawski and M. Gazdzicki
for providing us the experimental data from the NA61/SHINE Collaboration.
Furthermore, we acknowledge support by the Deutsche Forschungsgemeinschaft 
(DFG, German Research Foundation): grant BR 4000/7-1,  by the Russian Science 
Foundation grant 19-42-04101 and   by the GSI-IN2P3 agreement under contract number 13-70.
Also we thank the COST Action THOR, CA15213.
We acknowledge funding from the European Union’s Horizon 2020 research and 
innovation program STRONG-2020 under grant agreement No 824093. 
I.G. acknowledges support by the DFG through the grant CRC-TR 211 
'Strong-interaction matter under extreme conditions' - Project number 
315477589 - TRR 211 and by HGS-HIRe for FAIR.
The computational resources have been provided by the LOEWE-Center 
for Scientific Computing and the "Green Cube" at GSI.




\end{document}